\documentclass[journal]{IEEEtran}
\ifCLASSINFOpdf
   \usepackage[pdftex]{graphicx}
   \graphicspath{{../pdf/}{../jpeg/}}
   \DeclareGraphicsExtensions{.pdf,.jpeg,.png}
\else
   \usepackage[dvips]{graphicx}
   \graphicspath{{../eps/}}
   \DeclareGraphicsExtensions{.eps}
\fi
\usepackage{eurosym}

\usepackage{amssymb,amsfonts}
\usepackage{algorithm,algorithmic}

\usepackage{graphicx}
\usepackage{textcomp}
\usepackage{xcolor}
\def\BibTeX{{\rm B\kern-.05em{\sc i\kern-.025em b}\kern-.08em
    T\kern-.1667em\lower.7ex\hbox{E}\kern-.125emX}}
        
\usepackage{amsfonts}
\usepackage{amsthm}

\newtheorem{lemma}{Lemma}
\newtheorem{definition}{Definition}
\newtheorem{example}{Example}

\theoremstyle{definition}

\usepackage{amsmath}

\newcommand{\ALOOP}[1]{\ALC@it\algorithmicloop\ #1%
  \begin{ALC@loop}}
\newcommand{\ENDALOOP}{\end{ALC@loop}\ALC@it\algorithmicendloop}

\usepackage{url,multirow}

\begin{document}
%
\title{Security-Aware Access Model for Data-Driven EHR System}
%
%
%

\author{
\IEEEauthorblockN{Ngoc Hong Tran$^1$, Thien-An Nguyen-Ngoc$^1$, Nhien-An Le-Khac$^2$, M-Tahar Kechadi$^{1,2}$}\\
\IEEEauthorblockA{$^1$The Insight Centre for Data Analytics\\
$^2$School of Computer Science\\
\textit{University College Dublin, Belfield, Ireland} \\
ngoc.tran@ucd.ie, thien-an.nguyen@ucdconnect.ie, nhien-an.le-khac@ucd.ie, tahar.kechadi@ucd.ie}
}

\maketitle

\begin{abstract}
Digital Healthcare systems are very popular lately, as they provide a variety of helpful
  means  to  monitor people's  health  state  as well  as  to  protect people  against  an
  unexpected  health  situation.   These  systems   contain  a  huge  amount  of  personal
  information in a form of electronic health  records that are not allowed to be disclosed
  to unauthorised users.  Hence, health data  and information need to be protected against
  attacks and  thefts. In  this paper,  we propose a  secure distributed  architecture for
  healthcare data  storage and  analysis.  It  uses a novel  security model  to rigorously
  control permissions of accessing sensitive data in the system, as well as to protect the
  transmitted data between distributed system servers and nodes.  The model also satisfies
  the NIST  security requirements. Thorough  experimental results  show that the  model is
  very promising.
\end{abstract}


%
\IEEEpeerreviewmaketitle

\section{Introduction}
\label{sec:Int}
E-Healthcare systems are very popular in recent  years, as it becomes a matter of urgency
to manage and control patients' data. This data can be used to support them in bad health
and  emergency situations,  to  evaluate practitioner  performance, or  to  operate as  a
health-caring  consultant\footnote{How  Health  Care  Analytics  Improves  Patient  Care,
https://health\-informatics.uic.edu/blog/how-health-care-analytics-improves-patient-care/}
as well. Hence,  users find the most  of healthcare applications and  devices helpful. In
order to  predict and  support patients  well-being, their data  should be  collected and
 stored in a form  of electronic health records (EHR). However, as  patients' data is very
 sensitive and must  be protected against leakage  and attacks, the patients  do not trust
 e-healthcare  systems.  For instance,  Orange  telecommunication  provider declared  that
 $78\%$   of   users   did   not   trust   companies   in   the   way   their   data   was
 used.\footnote{Rethinking  Personal   Data:  Trust  and  Context   in  User-Centred  Data
 Ecosystems,
 http://www3.weforum.org/docs/WEF\_RethinkingPersonal\-Data\_TrustandContext\_Report\_2014.pdf}.
 Therefore, users'  trust has been  raised, as their personal  data must be  protected and
 processed   in   a    secret   way   against   possibilities    of   being   unexpectedly
 disclosed. Moreover, increasing user's trust  will surely guarantee the sustainability of
 the potential data-driven economy.

 To  protect patients'  data against  any abusive  use or  attacks, several  international
 regulations \cite{Rule15} have  been issued, such as the US  Health Insurance Portability
 and  Accountability Act  (HIPAA),  the  Health Information  Technology  for Economic  and
 Clinical  Health  (HITECH),  the  European  Union's  Data  Protection  Directive,  the
 Australian  Privacy Act  and Japan's  Personal Information  Protection Act  (PIPA). These
 regulations set policies on data access to the health information in the majority regions
 of  the  world.   However,  they  do  not protect  personal  data  directly  against  all
 vulnerabilities.  Therefore,  the risks of  leaking patient's personal  information still
 exist.
 

Marci
et al. in \cite{Marci2006} analysed the potential  issues of security and privacy that can
affect healthcare systems, such  as access to data, how and when  data is stored, security
of  data  transfer,  data  analysis  rights,  and  the  governing  policies.  Among  them,
\cite{Liu2015} \cite{Ahmad2016} \cite{Huang2017} protected the data during the storage and
access processes,  while \cite{Boon2009} \cite{Marcos2015} \cite{Guo2012}  proposed secure
solutions for data when it is transmitted through the network.

In \cite{ThienAnetal},  a large-scale healthcare system  was built to collect  and analyse
the health information about obesity in children.  The data is collected from patients and
voluntary students in different schools, cities,  countries and regions in Europe. Besides
the data  analytics, the  question is how  to protect this  big data  against unauthorised
users. The  proposed security  protocol is  similar to  \cite{ThienAnetal}, but  with more
efficient policies and mechanisms. In this paper, we propose a distributed architecture to
manage healthcare data more autonomously and process data queries more rigorously and with
high security. 

The  remainder  of the  paper  is  organised  as follows.  Section  \ref{sec:RelatedWorks}
presents  the  related   works  along  with  their  drawbacks   and  trade-offs.  Sections
\ref{sec:EHR}  and  \ref{sec:Archi}  describe  the   adopted  healthcare  system  and  its
architecture, respectively.  Section \ref{sec:UserAUTH} describes user  authentication and
authorisation mechanisms. Section \ref{sec:CAE}  presents controller authentication
and  its authorisation  methods.  The secure  data transmission  is  presented in  Section
\ref{sec:SDT}. Section  \ref{sec:SecEval} discusses the  evaluation of the  proposed model
against   the   security  attacks.   Experimental   results   are  reported   in   Section
\ref{sec:Experi}. We conclude in Section \ref{sec:Concl}.  

\section{Related Work}
\label{sec:RelatedWorks}

Health care systems have been under development for  quite a while. Their main goal is to
 reduce the administrative  cost, operational time, and to improve  the work efficiency in
 medical  related works.  In order  to  retrieve patients'  data more  effectively and  to
 provide richer data  sets for health measurements, authors in  \cite{Schatz15} proposed a
 database using mobile health monitors. They can collect a huge amount of data for further
 processes in predicting probabilities of chronic diseases, for instance. Whereas, authors
 in \cite{Yu15} built a healthcare system  with data analytics capabilities for diagnosing
 diseases early from symptoms provided by  patients. The authors exploited the data mining
 technique, namely  Naive Bayes, to detect  diseases from given symptoms.  These works did
 not focus on the privacy and security issues.

 Privacy and  security issues in the  healthcare information management systems  have been
 emphasised  for  years. To  address  privacy  risks  in sharing  healthcare  information,
 \cite{Ostherr17} outlined how sensitive this data is  and any disclosure of such data may
 have  devastating consequences  on  patients.  Furthermore,  Kotz  et al.   \cite{Kotz16}
 discussed  several issues  from policy,  regulation to  technologies, such  as anonymity,
 hash,  homomorphic cryptography,  and differential  privacy. However,  the point  is that
 these technologies  cannot guarantee  a complete privacy.   There is still  a need  for a
 security-aware model  that combines several  security and privacy technologies  to ensure
 data confidentiality and user privacy in healthcare systems.

 Some works had  different approaches in designing the security-aware  system framework in
 protecting user privacy or detecting the system risks  to get back to defend users. As in
 \cite{Dubo2017}, the authors proposed a  privacy-aware framework for managing and sharing
 electronic  medical  record  data for  the  cancer  patient  care  based on  block  chain
 technology. \cite{Tahar2006} proposed a secure  framework for a distributed network based
 on grid technique and  encryption techniques in controlling data access  through a set of
 policies. Another work \cite{Tahar2015} presented  a security-aware model to forensically
 visualise the evidence and attack scenario in a computer system from network log files so
 that the system operation can avoid such attacks. Authors in \cite{Tahar2016} proposed an
 interesting  security-aware   framework  to   protect  the   system  from   the  internal
 attackers. In this work, authors protect the  system resources by creating walls, each of
 which owns different security properties, between  users and the system resources to make
 users unable  to recognise  those resources.  Another work by  Liu et  al. \cite{Liu2015}
 built up the access control applied Role-based  Access Control (RBAC) to protect the user
 privacy in EHRs stored in the database from any outside access. This work can protect the
 data from  external accesses, however,  the secure data  transmission has not  been dealt
 with.

 For cryptographic solutions, the authors in \cite{Rao15} discussed possible solutions for
 security and privacy issues of big data heathcare systems.  They studied a solution based
 on  de-identification, using  data-centric approach  that  allows data  to be  decrypted,
 untokened  or unmasked  by  only authorised  objects  or access  control.  Hence, a  more
 detailed solution  is needed for a  distributed system and  protect it from all  risks of
 disclosing  the   patients'  information.   Authors  in   \cite{Blobel2016}  adopted  the
 de-identification and  proper ID management  techniques to hide important  information in
 the  database.  Whereas,  in  \cite{ThienAnetal},  the authors  focused  on  the  privacy
 preservation by applying de-identification and anonymity techniques to prevent the health
 information to  be disclosed  from the  data storage.  Another  approach in  applying the
 encryption technique was proposed by Miao  et al. \cite{Miao2016}. They designed a secure
 model  for searching  the  health information  from the  encrypted  database by  applying
 encryption   algorithm.   In   addition,   some   other   works   as   \cite{Barbara2013}
 \cite{Barbara2014}  \cite{Barbara2015}  applied  the homomorphic  encryption  and  secure
 comparison algorithms  to make the  collaboration among parties  in the system  unable to
 violate  the data  of the  other sides.  Some other  research works  focus on  leveraging
 distributed ledger technology  in protecting the health  information.  In \cite{dwi2019},
 the authors preserved  the privacy of medical data stored  and transmitted among Internet
 of  Things (IoT)  devices  by adopting  block chain  approach  and symmetric  lightweight
 cryptographic    algorithm   in    the   distributed    network.   In    \cite{Zhang2010}
 \cite{Shafer2010}, the authors focused on  securing data transmission by applying TLS/SSL
 solutions and encryption  algorithms such as AES  and RSA, but the  data storage security
 has not been dealt with in their solutions.

 From another  approach view,  in this paper,  we propose a  solution for  preserving user
 privacy and data  security in both data  transmission and storage for  an effective dealt
 with the big health data. More specifically, we propose a distributed model for executing
 methods of protecting  data against attacks such as replay  attack, eavesdropping attack,
 and unauthorised spy.  In order to set  such objectives, we exploit  the current powerful
 techniques,  such as  Kerberos,  LDAP,  Active Directory,  SSL/TLS,  etc., combined  with
 different cryptographic algorithms AES CBC/GCM, RSA, DES, SHA-2, ECDSA, etc.

\section{Data Access Security Settings}
\label{sec:EHR}
In the  following we define the  end-user role categorisation in  detail, their respective
permissions to data access, and other system components' permissions. 

\subsection{User Role Categorisation}
\label{sec:user_role}
\begin{figure}[h]
\centering
  \includegraphics[scale=0.3]{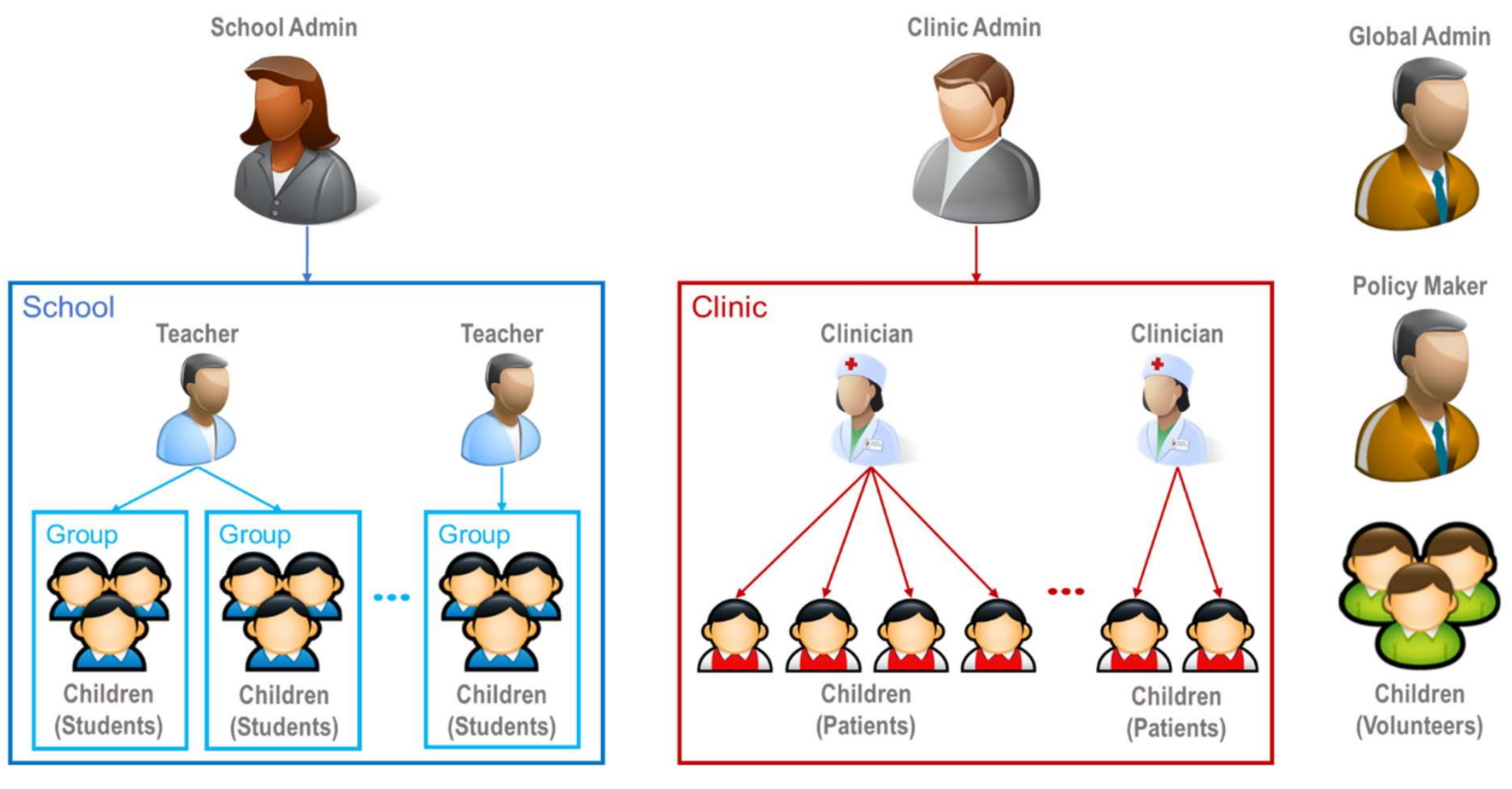}
  \caption{EHS User Categorisation.}
  \label{fig:EHClassification}
\end{figure}

In our system,  we define different classes  of users depending on their  roles, and grant
different permissions for  each class role. Each  user has a pair of  {\it username}, {\it
  password}, and roles  in the system. A user  can have more than one role  in the system.
Figure \ref{fig:EHClassification} shows  the relationship among user roles  in the system.
More specifically, there are three classes of  users: {\it school}, {\it clinic}, and {\it
  admin}.   For  each  class,  different  access rights  were  defined  to  reflect  their
activities  and interactions  with the  system.  The  {\it admin}  user manages  users and
classes. The roles are defined in the following:
\begin{itemize}
  \item Each class  {\it school} has a  {\it school admin} for managing  the school users,
        who may have  different roles. The school  user roles are {\it  teacher}, and {\it
        students}. In the school, there are  many {\it group}s of \textit{\it students}. A
        group is managed by a teacher. 
  \item Each class  {\it clinic} has a  {\it clinic admin} for managing  clinic users. The
        clinic roles are {\it clinician} and {\it patients}. A clinician manages a number
        of patients in the clinic.
  \item  The {\it  admin}  class includes  the core system  administrators, admin  outside
        schools and clinic admin staff. In this  class, a system admin, called {\it global
          admin},  manages  all the  users  of  the whole  system,  a  {\it policy  maker}
        identifies  conditions  associated  to  childhood obesity  and  designs  effective
        policies, which are then applied to hospitals, schools, communities, etc.
\end{itemize}

Moreover, the  storage system has  two major databases, which  are MongoDB\footnote{https:
  //www.mongodb.com/} and  Cassandra\footnote{http://cassandra.apache.org}.  The Cassandra
database stores the time-series data collected from users' wearable devices, such as smart
watch and smart  sensors, and smart phone,  whereas, the MongoDB database  stores the most
sensitive data of  the system.  These include  user management and other  private data. In
this  work, we  apply Role-based  Access Control  (RBAC) technique  to efficiently  manage
system's users based on  their roles. The admin information are  stored in  MongoDB. These
include the user's registration information and  other information required by the system,
such as {\it username, password, roles and global ID number, etc}. Depending on the roles,
a user may have  other administrative details such as {\it sub-role,  group ID, school ID,
  clinic ID, and  supervisor ID}.  All collected  and derived data of  a participant, that
was stored in either MongoDB or Cassandra is  identified with a global ID. In other words,
a unique global ID is assigned to each participant in the system to reference any piece of
their data in the  BigO system.  Users are granted access permissions  to the system based
on their role. For  example, the clinician can access the data of  their patients only via
clinic portal, the teacher can access data  of their students only via school portal. More
details of the user verification process are presented in Section \ref{sec:UserAUTH}.

\subsection{Permission Settings}
The system  consists of two  types of components:  controllers and modules.   A controller
includes a set of services provided to  end-users or to the other modules. Each controller
has  limited privileges  for  providing access  to the  back-end.   Therefore, all  access
permissions  requested  by  the  controllers  are checked  based  on  a  technique  called
Discretionary Access Control  (DAC). Each controller is evaluated for  its own permissions
for each collection, and it may have the permissions to {\it read}  ({\bf R}) a data, {\it
  write} ({\bf W})  a data or both {\it  read} and {\it write} ({\bf RW}).  With a reading
permission, for instance, a controller can make  {\it select} or {\it find} request to the
database ({\it select} query for MongoDB and {\it find} query for Cassandra). For writing,
a controller can send an {\it insert/update/remove}  or {\it modify} request to any if the
two databases.

\begin{figure}[h]
\centering
  \includegraphics[width=6cm, height=5cm]{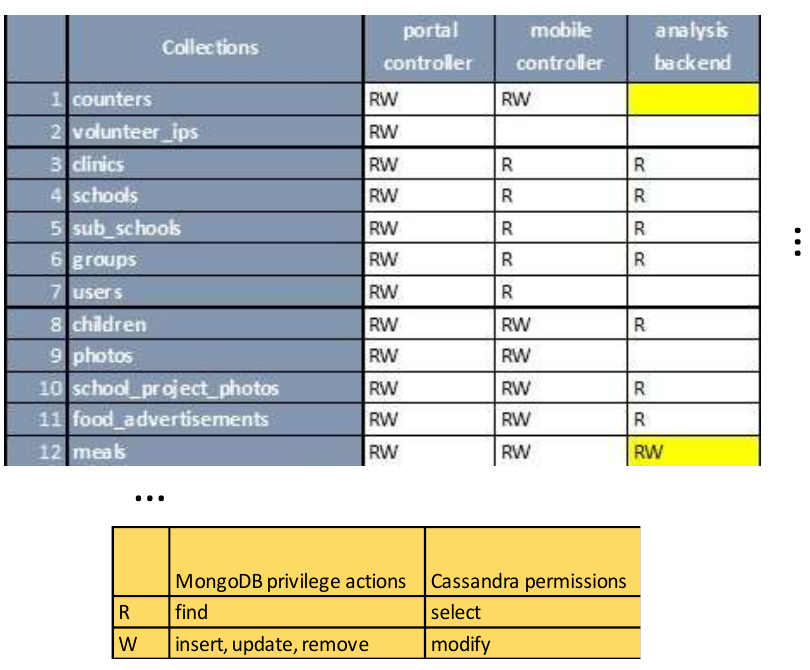}
  \caption{DAC Table of Controllers vs Collections}
  \label{fig:DAC}
\end{figure}

In the DAC Table (see Figure \ref{fig:DAC}), a number of controllers are provided with the
reading and writing  permissions with the respective data, that  is, collections/tables in
the databases. The columns in DAC table are the system controllers, and the rows represent
collections/tables. For instance, the {\it portal controller} has {\bf  RW} permissions on
 all the  collections. A detailed  description of  the controller verification  process is
 presented in Section \ref{sec:controller_authorization}. 

\section{Security-Aware Architecture}
\label{sec:Archi}
We adopted the MongoDB and Cassandra database  management systems to leverage the power of
each  of them,  especially, a  combination  of the  both  can exploit  fully their  strong
characteristics. Specifically,  Cassandra manages  only user data  crawled from  the smart
mobile devices, and  its very fast writing  queries. Whereas, MongoDB manages  the rest of
user information  (e.g., users, clinics,  schools, timelines, regions,  statistics, etc.),
and   it  supports   a   good   querying  performance   for   its  feature  \textit{index
  settings}. Moreover,  MongoDB supports  authentication and authorisation  using Kerberos
\cite{Kerberos} and  LDAP \cite{LDAP}, as  well as AES encryption\footnote{Announcing the Advanced Encryption Standard (AES), NIST, FIPS 197, 2001}  at-rest and
in-flight  modes. Cassandra  supports authentication  and authorisation  using LDAP,  data
transmission using Transparent  Data Encryption (TDE)\footnote{Oracle Advanced Security Transparent Data Encryption Best Practices, https://www.oracle.com/technetwork/database/security/twp-transparent-data-encryption-bes-130696.pdf}.  The security techniques
supported  by MongoDB  and  Cassandra  fulfil the  security  requirements against  current attacks.\footnote{Security Requirements for Cryptographic Modules, Federal Information Processing Standards Publication, NIST, FIPS 142, 2001} 

\begin{figure}[h]
  \includegraphics[width=8.5cm, height=7.2cm]{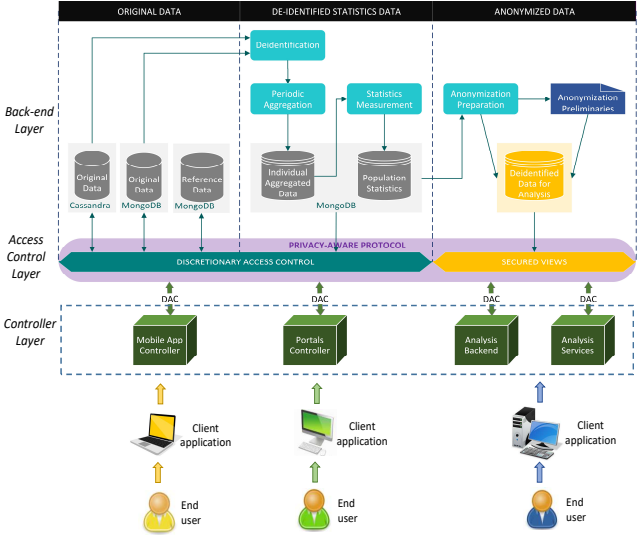}
  \caption{Security-aware $3 \times 3$ matrix model}
  \label{fig:architecture}
\end{figure}

We adopted a  security design architecture at  every layer and between  layers (see Figure
\ref{fig:architecture}). The layers are {\it controller layer}, {\it access control layer}
and  {\it   back-end  layer}.   Each  layer   verifies  the   permissions  of   the  lower
layer. Therefore, it is more strict to check  the data access from the bottom layer to top
layer. Within  each layer one needs  to deal with the  security of the following  types of
data  {\it  original data},  {\it  de-identified  statistics  data}, and  {\it  anonymised
  data}. The security level  increases from original to anonymised data,  as the data gets
de-identified, anonymised and perturbed respectively.

Consider  the controller  layer, a  controller contains  a set  of services  providing the
end-users access from the external network or  from another component from the system. The
{\it  Mobile App  Controller} provides  a set  of interfaces  for mobile  users to  access
services such as  inputting user's information about meals, activities,  etc. These mobile
data values are saved in the mobile  device memory before their transfer to Cassandra. The
{\it  Portal controller}  supports a  set of  interfaces that  can be  used to  access the
supported  web services,  such  as  {\it Clinical  Analysis  Services}  for measuring  the
activity rates, predicting the effect of  users' activities, etc. This data is transmitted
to MongoDB or the related {\it Analysis Services}. Any controller access that is requested
by the end-users are evaluated using RBAC as mentioned in Section \ref{sec:user_role}. The
two  other controllers;  {\it Web  Portal}  and {\it  Analysis Services}  access the  data
through a Secured View. 

For the  access control layer, there  are two investigations for  evaluating its requests,
that is, {\it DAC} and {\it Secured  View}. These parts are executed by the authentication
server.   Only Mobile  App Controller  and Portal  Controller can  access DAC,  while only
Analysis Backend and Analysis Services can access Secured View. DAC determines weather the
requesting  controller  is eligible  to  access  the data  in  the  back-end layer  before
performing its  request. The  Secured View controls  and restricts access  to the  data by
encouraging the controllers  to use de-identified and anonymised data  in order to protect
data from possible vulnerabilities of leaking the personal information.

At  the back-end  layer, the  database  server protects  data  one more  time against  the
unauthorised controllers and end-users by determining weather that controller and user can
access such requested data.   The {\it Original Data} is raw data and  it is readable. The
{\it Identified Statistics  Data} contains data excluding the identity  information of the
users. The {\it Anonymised Data} stores data  without leaving any trace of an inference to
the user identity. 

Furthermore, components  of $3\times  3$ matrix  are located  in a  physically distributed
servers.  In particular, databases are stored in different servers, i.e., original MongoDB
data, original Cassandra  data, and reference data  of original data column  are stored on
the three separate servers. The individual  aggregated data and population statistics data
are stored  on one server, and  the de-identified data  for analysis is stored  on another
server. The distributed database storage plays a key role in improving data management and
privacy against the linkability attack.

\begin{figure}[h]
\centering
  \includegraphics[width=2.5cm, height=2.5cm]{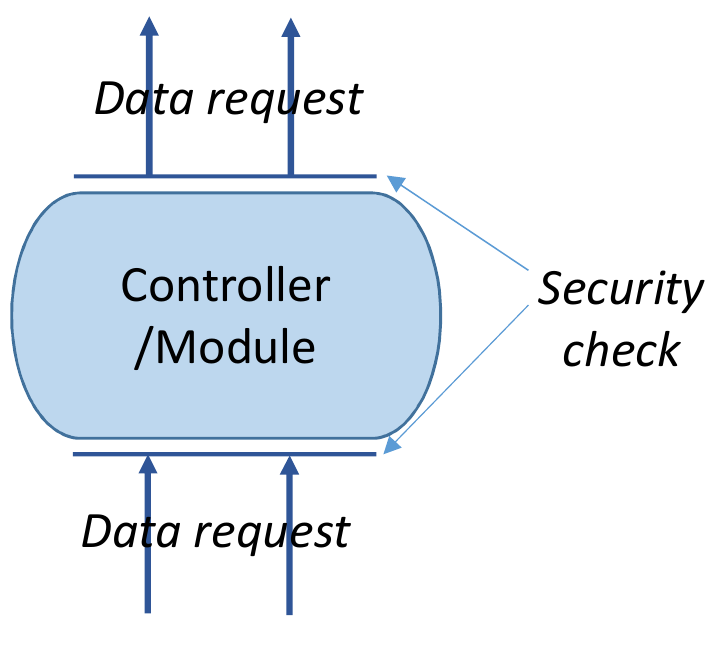}
  \caption{Double security check at each component and module.}
  \label{fig:2_secure_check}
\end{figure}

Moreover, each component shown in Figure \ref{fig:architecture} is independent. Therefore,
the process  of receiving a request  from a component  and transferring the result  to the
next component can cause a privilege violation,  if the security between components is not
enforced. Hence,  the data flow  between components should  be rigorously verified  on its
path at  each component hop.   Each component identifies  the client and  evaluates its
permissions before  processing the received request.  Then it transfers the  result to the
next  component.   The   next  component  again  evaluates   the  client's  permissions
(c.f.  Figure \ref{fig:2_secure_check})  and so  on  until the  last hop  is reached.  For
example,  in Figure  \ref{fig:architecture} only  mobile  app and  portal controllers  can
access directly to Role-Based Access Control (RBAC) component of the access control layer,
to query the  original and statistical datasets, and only  Web Applications Controller and
Analysis Services can access Secured Views of the access control layer. The RBAC component
stores mappings  of which  controller/user can  access which  database, while  the secured
views store anonymised data that remove the Quasi-identifiable (QI) data \cite{QI}. 

\section{End-user Privilege Check}
\label{sec:UserAUTH}
A   request  transaction   can  be   described   a  security-aware   $3\times  3$   matrix
model. Basically,  the controller  layer implements  a set  of REST  APIs to  support user
queries to database. An end-user who wants to  access a database needs to log into a given
controller.  This is  the first  shield  of the  model  in protecting  the database.  Each
end-user $u$  has their own  credential (i.e., user name  $un_u$ and password  $p_u$). The
user can have one  or more roles, denoted as $lrole_u$, in the  system (e.g., a teacher, a
clinician, a school  administrator, a clinic administrator, etc.), a  unique ID $ID_u$ and
their organisation  ID $orgID_u$ (e.g., school  ID, clinic ID, etc.).  To authenticate and
authorise a user, the {\it Role-Based Access Control} (RBAC) technique \cite{RBAC} is used
to  handle the  data  access  permission of  controllers.  The role  of  a  user or  their
organisation is checked for any data access. The details are given in the following.

\subsection{Token-based User Authentication}
\label{sec:tokenbasedUserAuthen} 
User authentication happens when a user sends  a log-in service request to the controller,
a long with their credential as an  input. The user credentials are formally formulated as
follows.

\begin{definition}[User Credential]
  \label{def:userCredential}
  Let  $u$  be a  user  of  the system.  Let  $(un_u,p_u)$  be a  pair  of  user name  and
  password. Let  $ID_u$, $lrole_u$, $org_u$ be  $u$'s ID, role list,  and organisation ID,
  respectively. The user credential $Cr_u$ can be defined as:
  $$Cr_u = [ID_u, (un_u, p_u), lrole_u, orgID_u]$$
\end{definition}

All passwords are  hashed before being stored  in the database. Based  on this credential,
the controller authenticates the user in  collaboration with a security server. To provide
the user  a legal authentication certification  to use for their  subsequent requests, the
controller gets  a JASON Web  Token (JWT)  generated by the  security server. This  JWT is
created using the  RSA algorithm with 512-bit key  size. The JWT is expired  after a $120$
minutes.  Therefore, the user has  to send a request to the server to  get a new JWT. With
the new JWT, the process can continue. More formally, the JWT is formulated as follows.

\begin{lemma}[JWT Token]
  \label{lemma:JWT}
  Let $u$, $Cr_u$  be a user and  their credential. $JWT_{(Cr_u,t)}$ is JWT  token at time
  $t$ for the user $u$ with credential $Cr_u$. It is defined as follows:
  
  \begin{equation}
  \label{equ:jwt}
  \begin{split}
   JWT_{Cr_u} =& [RSA512(Header_{(Cr_u, t)})|\\
               & RSA512(Payload_{(Cr_u, t)})|\\
                & RSA512(Sign_{(Cr_u, t)})]\\
    with\ 
    Header_{(Cr_u,t)} &= \{alg_{jwt}|type_{jwt}\}\\
    Payload_{(Cr_u, t)} &= \{lrole_u | iss_u | ID_u |\\  & exp\_time_{(u, t)}| version_{u,t}| iat_u\}
   \end{split}
\end{equation}

  where 
  \begin{itemize}
     \item $'|'$ is a concatenation operator.
     \item $Header_{(u, t)}$ contains algorithm  $alg_{jwt}$ (e.g., RSA512) and token type
           $type_{jwt}$. 
     \item $Payload_{(u,  t)}$  is the  token  body  involving  a  user $u$'s  role  list
           $lrole_u$, the issuer $iss_u$ grants this  token for $u$, an identity $ID_u$ of
           $u$, an expiration time $exp\_time_{(u, t)}$ computed since time $t$ to use the 
           services, version of token, and the issuing time of token.
     \item $Sign_{(u, t)}$  is the signature of  the header, payload and  secret key using
           SHA-512.
  \end{itemize}
\end{lemma}

\begin{example}
  \label{ex:JWT}
  An example  of JWT  token, generated  from user  $u$'s credential,  is given  in Figures
  \ref{fig:endec_JWT}. Figure \ref{fig:endec_JWT}a  presents a code of an encoded JWT, and
  Figure \ref{fig:endec_JWT}b shows a structure of post-decoded JWT. 
  \begin{figure}[h]
  \centering
    \includegraphics[width = 8cm, height = 5.5cm]{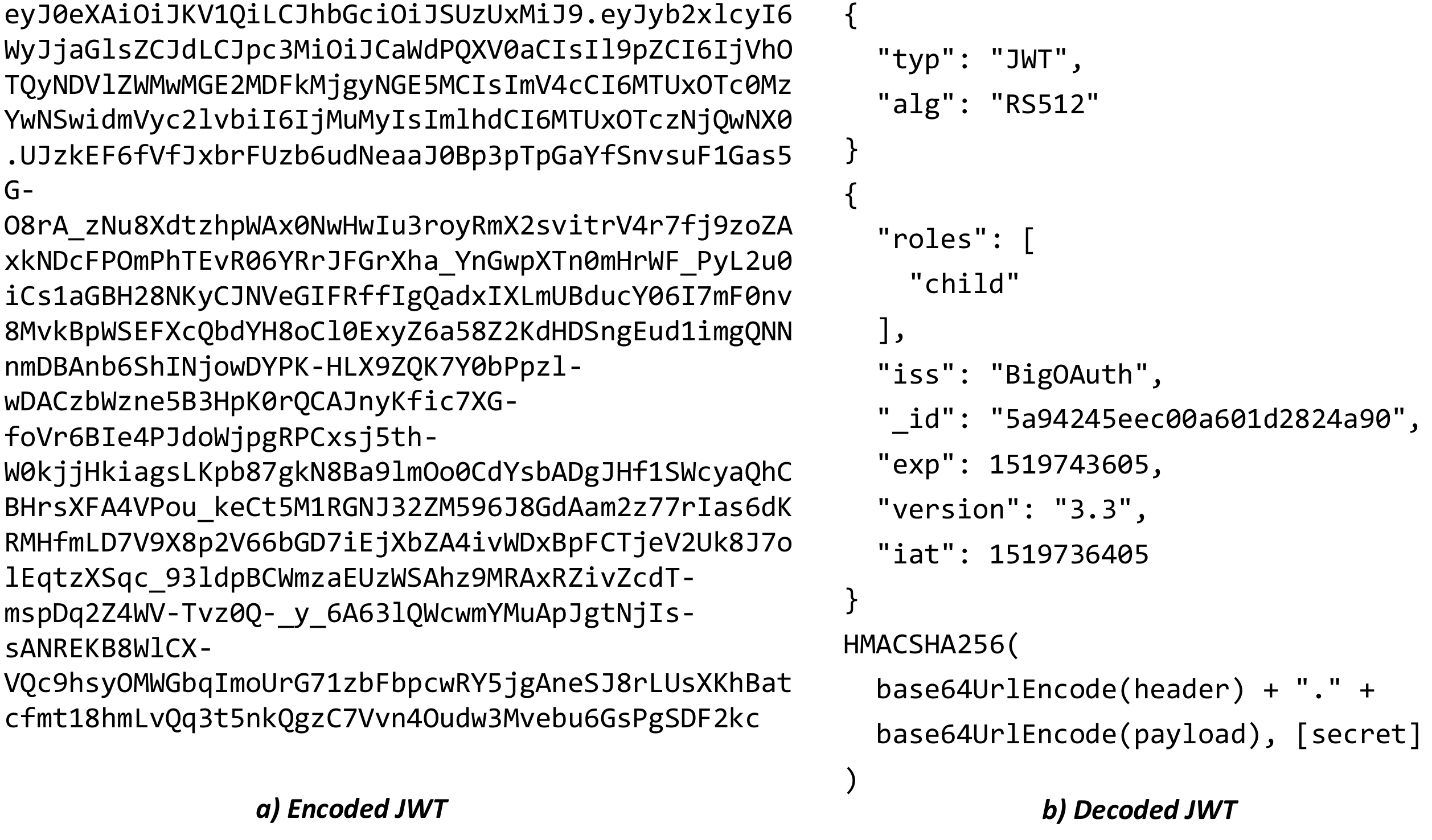}
    \caption{An example of an encoded and decoded JWT}\label{fig:endec_JWT}
  \end{figure}
\end{example}

\begin{lemma}[Unique User Token]
  \label{lemma:unique_User_Token}
    Let  $t$  denote  the  time  and  $u$  the  user  with  credential  $Cr_u$. The  token
  $JWT_{(Cr_u,t)}$ is unique for user $u$ within a period of time $t+\Delta$.
\end{lemma}

\begin{proof} 
  JWT token is  based on the current date time  $t$ of the server. The time  $t$ is unique
  and it is  obtained automatically from the  server.  The user name  and their credential
  are unique. Only one  hash value using SHA512 appended into the JWT  token as the hashed
  content is unique  with the unique secret  key and the unique  issuing time.  Therefore,
  there is only one $JWT_{Cr_u}$ to be generated according to one $Cr_u$.
\end{proof}

The Lemma \ref{lemma:unique_User_Token} guarantees that a legal user can always access the
controller's services in an  interval of time $\Delta$. As the same  user credential and a
requested time,  the generated token  is unique, hence,  the controller can  recognise the
user identity through their token uniquely.

After successful  authentication, the  end-user $u$ can  access the REST  APIs of  a given
controller  using their $JWT_{(Cr_u,  t)}$.   Whenever a  user  $u$  requests a  service,
$JWT_{(Cr_u,  t)}$  is  embedded  into  that  request.  The  called  REST  API  checks  if
$JWT_{(Cr_u,  t)}$ is  valid, and  the user's  roles in  the $JWT_{(Cr_u,  t)}$ cover  the
permission of using the called request.

\subsection{Collection-based User Authorisation}
\label{sec:CollectionbasedUserAuthorize}
Database  stores   user  information  using   a  specific  data  structure,   called  {\it
  collection}. A collection is a list of documents and each document keeps the information
of  one user.   This  storage  method does  not  check if  all  documents  have a  uniform
format. To authorise a user, the controller  needs to seek their relevant information in a
collection called "USER" of the database where all users'  information are saved. Then, it
checks the user's permission on the requested  data by checking their credential and their
administrative  information. Specifically, a user  permission is  well authenticated if it 
satisfies  two  conditions, that is, (C1) only the administrative  staff  can  access  the 
information of users they manage (see Definition \ref{def:legal_stewardship}),  and (C2) a
staff  member  can only read or write the information of a user in their organisation (see 
Definition \ref{def:solid_membership}). 

\begin{definition}[Legal stewardship]
  \label{def:legal_stewardship} 
Let $u$ and $v$  be the data user and owner, respectively. Let  $lrole_u$ be the role list
of $u$.  Let $orgID_v$  be the organisation ID of $v$. A  Legal Stewardship is established
between $u$ and $v$ iff:
$$\exists i | ((role_u[i] \in lrole_u) \wedge (role_u[i] \in admin(orgID_v)))$$
where $admin()$  is the  function returning a  list of the  administrative roles  of $v$'s
organisation. 
\end{definition}

\begin{definition}[Solid membership]
  \label{def:solid_membership} 
Let $u$, $v$ be the data user and owner, respectively. Let $orgID_u$, $orgID_v$ be the IDs
of the organisations of $u$ and $v$,  respectively. A Solid Membership between $u$ and $v$
exists iff:
$$orgID_u = orgID_v$$
\end{definition}


\begin{lemma}[User Authorisation]
User $u$ can access the data of a patient $v$ if and only if user $u$'s permission satisfies
the two conditions (C1) and (C2) as in Definitions \ref{def:legal_stewardship} and 
\ref{def:solid_membership}, respectively.
\end{lemma}

\begin{example}
Let consider the case where  the user is a clinician $A$ of Clinic  $X$. $A$ makes a query
on the  user $B$'s information  from the  database. Moreover, $B$  is a patient  of Clinic
$X$. Therefore, $C1$  and $C2$ are satisfied.  Then, clinician $A$ is allowed  to query on
$B$'s information from the database. 
\end{example}

The  authorisation procedure  is  described in  Algorithm \ref{alg:authorizeUser},  namely
\textit{authorizeUserPermission()}, to check the user $u$'s permission on some data of $v$
which $u$ aims to access. Basically, as mentioned in Definitions \ref{def:legal_stewardship},   \ref{def:solid_membership},   the  $u$'s   information,
involving $u$'s role list $lrole_u$ and $u$'s organisation ID $orgID_u$, are achieved from
their credential (cf. Definition \ref{def:userCredential}).  To satisfy both lemmas above,
$v$'s role and organisation ID need to be obtained. $v$'s information are queried from the
relevant collections through $v$'s ID. 

\begin{algorithm}
  \caption{\textit{authorizeUserPermission()}}
  \label{alg:authorizeUser}
  \begin{footnotesize}
    \begin{algorithmic}[1]
    \REQUIRE $lrole_u$, $orgID_u$, $ID_v$
    \ENSURE $'true'|'false'$
    
    \STATE $orgID_v = findOrginCol("USER", ID_v)$;\label{alg1:lineSeekVOrgID}
   	\STATE $lroleAdminOfOrg_v = findAdminOfVOrg(OrgID_v)$;\label{alg1:lineSeekListAdminRoleOfV}
     
    \STATE $isRightRole = false$;\label{alg1:lineSetVar1False}
    \STATE $isSameOrg = false$;\label{alg1:lineSetVar2False}
    
    \FOR{$(\forall{role_i} \in lrole_u, i=0...n)$}\label{alg1:lineForI}
	    \FOR{$(\forall{role_j} \in lroleAdminOfOrg_v, j=0...m)$}\label{alg1:lineForJ}
			\IF{$(role_i == role_j)$}\label{alg1:lineCheck2Role}
	    		\STATE $isRightRole = true$;\label{alg1:lineSetSuccessCheckRole}
	    		\STATE $break$;\label{alg1:lineExitLoopI}
			\ENDIF
		\ENDFOR
		\IF{$(isRightRole == true)$}\label{alg1:lineIsRightRole}
			\STATE $break$;\label{alg1:lineExitLoopJ}
		\ELSE
			\STATE $return\ false$;\label{alg1:lineReturnFailedCheckRole}
		\ENDIF
    \ENDFOR
    
    \IF{$(orgID_u != orgID_v)$}\label{alg1:lineCheckOrg}
    	\STATE $isSameOrg = false$;\label{alg1:lineSetFailCheckOrg}
		\STATE $return\ false$;\label{alg1:lineReturnFailedCheckOrganization}
    \ENDIF

	\STATE $return\ true$;\label{alg1:lineReturnSuccess}

    \end{algorithmic}
\end{footnotesize}
\end{algorithm} 

Specifically, the  Algorithm \ref{alg:authorizeUser} inputs are  the list of roles  of the
user $u$,  their organisation  ID, and $v$'s  ID.  The algorithm  outputs a  boolean value
$"True"$ if the authorisation is successful, and  returns a boolean value $"False"$ if the
authorisation failed.  Initially, the necessary information  of $v$ are retrieved, such as
their organisation ID  $orgID_v$ (cf.  line \ref{alg1:lineSeekVOrgID}) based  on which the
list  of  administrative roles  in  $v$'s  organisation $lroleAdminOfOrg_v$  are  obtained
(cf. line \ref{alg1:lineSeekListAdminRoleOfV}). Moreover,  to check the two aforementioned
conditions $C1$ and $C2$, we initialise the two Boolean variables, i.e., $isRightRole$ and
$isSameOrg$.   $isRightRole$  contains the  result  of  $C1$ evaluation,  and  $isSameOrg$
contains the result  of $C2$ evaluation. Initially,  both are set to  $"False"$ (cf. lines
\ref{alg1:lineSetVar1False},   \ref{alg1:lineSetVar2False}).    Then,  $u$'s   role   list
$lrole_u$ is  examined (cf. line  \ref{alg1:lineForI}). The administrative roles  in $v$'s
organisation    $lroleAdminOfOrg_v$   are    investigated   simultaneously    (cf.    line
\ref{alg1:lineForJ}).  Each element  $role_i$ in $lrole_u$ is  compared to each
element $role_j$  in $lroleAdminOfOrg_v$.  If  there is  a case of  two matched
elements (cf. line \ref{alg1:lineCheck2Role}), $isRightRole$  is set to $"True"$ (cf. line
\ref{alg1:lineSetSuccessCheckRole}),   and  the   two  loops   are  stopped   (cf.   lines
\ref{alg1:lineExitLoopI},      \ref{alg1:lineIsRightRole},      \ref{alg1:lineExitLoopJ}).
Otherwise,     the    algorithm     stops    and     return    $"False"$     (cf.     line
\ref{alg1:lineReturnFailedCheckRole}).   In  the  case  of $C1$  is  successful,  function
$authorizeUser()$ continues to check $C2$ through the  two organisation IDs of $u$ and $v$
(i.e.,  $orgID_u$  and  $orgID_v$)  (cf.   line  \ref{alg1:lineCheckOrg}).   If  both  are
different,    the   variable    $isSameOrg$    is   set    to    $"False"$   (cf.     line
\ref{alg1:lineSetFailCheckOrg}), function $authorizeUserPermission$ is stopped and returns
$"False"$ (cf. line \ref{alg1:lineReturnFailedCheckOrganization}). In  the case of the two
conditions $C1$  and $C2$  are satisfied,  the authorisation  is successful,  and function
$authorizeUserPermission()$ returns $"True"$.

\section{Controller Authority Evaluation} 
\label{sec:CAE}
The  second  part of  the  model  protecting the  database  is  the access  control  layer
(cf.   Figure    \ref{fig:architecture})   which    manages   the   data    influence   by
controllers.  In the  proposed  architecture,  databases in  the  back-end  layer are  not
accessible  by all  controllers. Each  database  is restricted  for a  number of  specific
controllers.   The controller  that receives  user  request and  makes a  connection to  a
certain database  to get the data  is called \textit{requesting controller}.  The database
that  contains the  requested data  is called  \textit{requested database}.  There is  an
authentication and authorisation process between the requested database and the requesting
controller whenever they  handshake for a data request/response transaction.  In our work,
\textit{Role-Based Access Control} (RBAC) technique  \cite{RBAC} is deployed to handle the
data  access  permission  of  controllers.  RBAC  is  executed  based  on  leveraging  the
controller's identity.

\subsection{Single-Sign-On Controller Authentication}
\label{sec:Cont_Aut}
To identify the requesting controller, the  access control layer verifies the controller's
credential.   Each  controller is  set  with  a controller  name  and  a password  by  the
developer.  Its name must  be unique in the system. Moreover, the  controller has a secret
key denoted  as $K_c$,  generated from  controller's password by  using some  certain hash
function.

\begin{definition}[Controller Credential] 
  Let  $c, cn_c, p_c$  be  the  verified  controller,  controller's  name,  and  password,
  respectively.   Let $ID_c$  and  $K_c$  be the  controller's  identity  and public  key,
  respectively. Formally, the credential of a controller is defined as:
$$Cr_c = [(cn_c, p_c), ID_c, K_c]$$
\end{definition}

\begin{figure}[h]
\centering
  \includegraphics[width=7.5cm, height=5cm]{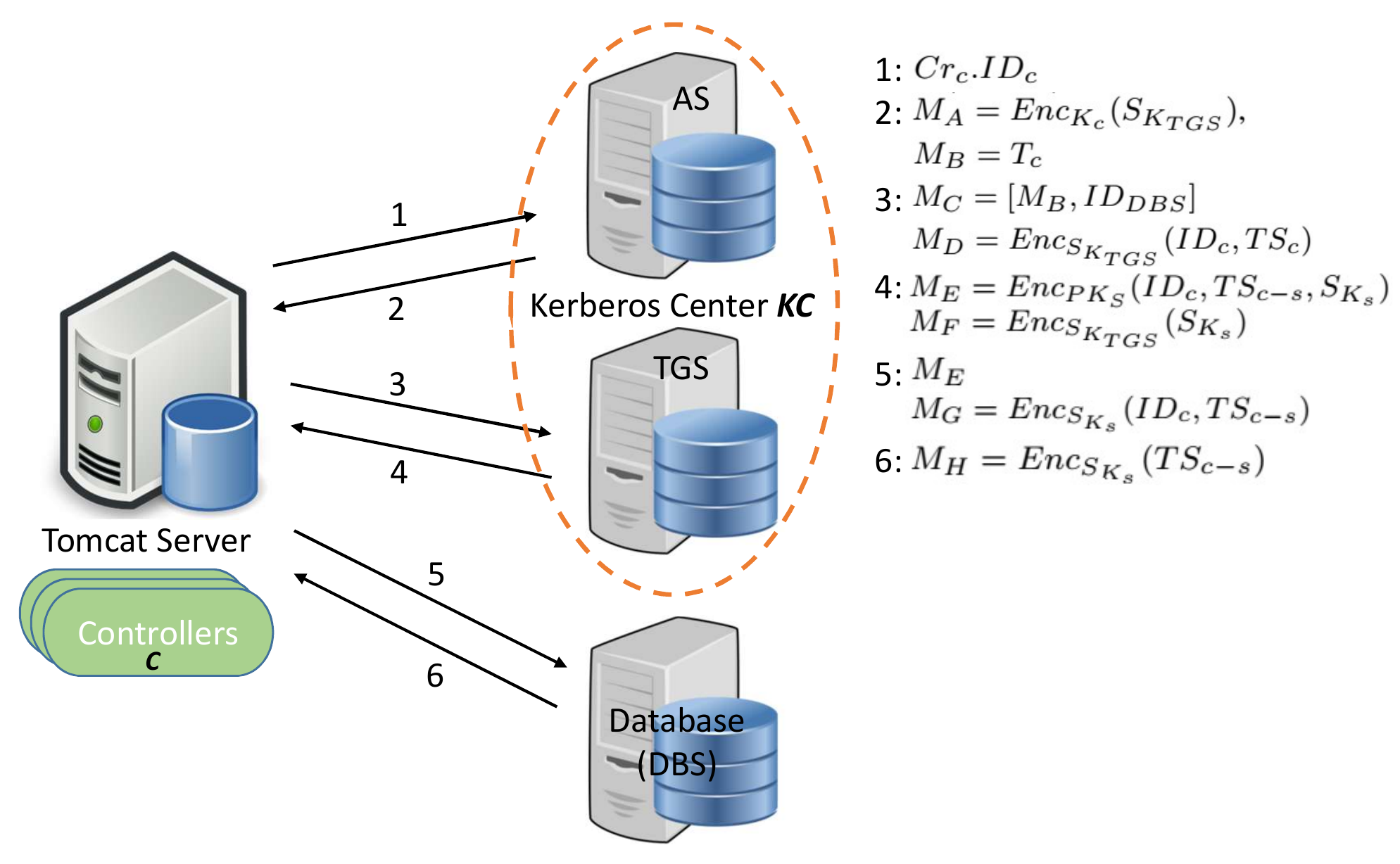}
  \caption{Keberos-based Controller Authentication}
  \label{fig:RAC}
\end{figure}

In order  to monitor several  users accessing system services,  a Kerberos Center  (KC) is
used  for   controlling  authentication  (cf.   Figure  \ref{fig:RAC}).  Kerberos   is  an
authentication protocol  between unknown  parties in  a secure  manner. $KC$  involves two
servers, that is, Authentication Server ($AS$) and Ticket Granting Service ($TGS$). In our
work, this  authentication protocol is  operated between  system controller and  $KC$. The
authentication is based on a ticket generated  by $AS$. Basically, the ticket is generated
using time  constraint that limits the  ticket's time-to-live ($TTL$), and  the controller
name is  saved in  the controller  credential. This  authentication ticket  guarantees the
``Single Sign On (SSO)''  property of the log-in system. It  indicates that the controller
needs  to  do authentication  with  $KC$  and then  it  uses  this ticket  for  subsequent
communications between the controller and the data. 

\begin{definition} [Service Ticket]
  \label{def:service_ticket} 
  Let $ID_c$, $TS_c$ be controller $c$'s ID  and a time stamp of the ticket, respectively.
  Let $PK_s$, $S_{K_s}$  be secret  key and  session key of  service $s$,  respectively. The
  service ticket of $c$, denoted $T_c$, can be generated, by an encryption algorithm:
  $$TK_c = Enc_{PK_s}(ID_c, TS_c, S_{K_s})$$
\end{definition}

Each ticket  has a time stamp  that protects it against  the replay attack. Once  the time
stamp  is reached,  a  new ticket  is  generated with  a new  time  stamp. Therefore,  the
adversary cannot  reuse the ticket  for another transaction.  Moreover, each ticket  has a
session  key  used for  transactions  between  the  service  and the  controller.  Tickets
generated  for  the  same  controller  are obviously  different  from  other  controllers'
tickets. In  addition, the encryption  of the ticket  is made with  the secret key  of the
service, hence, it can be decrypted by only the same service.

\begin{lemma}[Unique Ticket]
  \label{lemma:unique_Controller_Ticket}
  Let  $c,  TK_c$ be  the  verified  controller and  ticket  that  TGS generates  for  $c$
  (cf. Lemma \ref{def:service_ticket}). $TK_c$ is unique.
\end{lemma}

\begin{proof} 
Given a  controller $c$  and a service  $s$ $c$ wants  to access  the data service  $s$. A
ticket   is  generated   for   $c$;   $TK_c$.  $TK_c$   is  unique.  Because, from   Lemma
\ref{def:service_ticket}, a  ticket is an encryption  of $c$'s identity $ID_c$,  a $s$'s
shared key $SK_s$, and a time stamp $TS_c$, where
\begin{itemize}
   \item The controller identity $ID_c$ is unique.
   \item The secret key of service $s$, called $SK_s$, is unique in the system.
   \item The encryption  made from the two above elements with the secret key of $s$
         cannot be duplicated in the system.
\end{itemize}
Therefore, tickets of controllers are unique.
\end{proof}

Furthermore,  Kerberos  operates  in  cooperation   with LDAP  server.  All  controller's
information  and  their  permissions are  stored  in  the  active  directory of  the  LDAP
server.  For  each  entry  in  the  active  directory,  it  contains  username,  password,
organisation,  group, domain  name (i.e.,  a link  attached to  the organisation),  access
permissions, etc.\footnote{Access control list syntax; https://www-01.ibm.com/software/network/ directory/library/publications/ dmt/ref\_acls.htm}

The  authentication protocol  is depicted  in Figure  \ref{fig:RAC}.  The  details of  the
protocol are given  in Table \ref{tab:RAC}.  To start an  authentication, a controller $c$
connects to  $KC$ and sends its  ID ($ID_c$) to $AS$  (step 1). Then, $AS$  uses $ID_c$ to
lookup  $c$ in  its  master database  (step  2.1). If  $c$  is not  in  the database,  the
connection is rejected.  Otherwise, $AS$  continues to process the authentication request.
$AS$ creates a time stamp for $c$ ($TS_c$)  (step 2.2). $TS_c$ is used for checking if the
ticket  is still  valid, otherwise,  $TS_c$ needs  to be  renewed. $AS$  then generates  a
ticket, $T_c$,  for $c$ that  is an  encryption of $c$'s  identity (i.e., $ID_c$),  a time
stamp, ($TS_c$), and  a session key of $TGS$ ($S_{K_{TGS}}$)  (step 2.3). $S_{K_{TGS}}$ is
used by $c$ and $TGS$ in subsequent steps. The ticket $T_c$ is used for authenticating $c$
with $TGS$  before getting  the permission  to access database.  Then, $AS$  retrieves the
password  of $c$,  that is  $Cr_c.p_c$ (step  2.4) to  re-generate the  secret key  of $c$
($K_c$) (step  2.5). $AS$ creates two  messages; $M_A$ and $M_B$,  to be sent back  to $c$
(step  2.6).  $M_B=T_c$  and  $M_A$  is  the  encryption  of  the  session  key  of  $TGS$
($S_{K_{TGS}}$).  So, $M_A=Enc_{K_c}(S_{K_{TGS}})$.  $AS$ sends  $M_A, M_B$  to $c$  (step
3). $c$ generates the  secret key $K_c$ from its password $Cr_c.p_c$  (step 4.1). $c$ uses
$K_c$ to decrypt  $M_A$ and obtain $S_{K_{TGS}}$  (step 4.2). If $c$ fails  to process the
decryption,  it  indicates  that  $c$  cannot  access the  system  as  their  password  is
invalid. If  $c$ successfully  decrypts the  message, it creates  two messages;  $M_C$ and
$M_D$ (step 4.3). $c$ sends $M_C$ and $M_D$  to $TGS$ (step 5).  $M_C$ involves $M_B$, the
identity  of the  requested service  ($ID_{DBS}$) and  the requested  command is  executed
$Req_{DBS}$  (reading  $'R'$, writing  $'W'$,  searching  $'S'$ and/or  comparing  $'C'$),
whereas, $M_D$ is an  encryption of $c$'s identity ($ID_c$) and the  current time stamp of
$c$ ($TS_c$). $TGS$ decrypts $M_B$ to $M_C$ by its secret key $PK_{TGS}$ (step 6.1) to get
the shared key $S_{K_{TGS}}$ and the identity  of $c$ ($ID_c$).  $TGS$ decrypts $M_D$ with
the retrieved  key $S_{K_{TGS}}$  to get the  identity of $c$  ($ID_c$) (step  6.2). $TGS$
compares the two identities of $c$ obtained in  steps 6.1 and 6.2.  If they are equal, $c$
is  authenticated  to  $TGS$.   The  controller  authorisation  is  described  in  Section
\ref{sec:Cont_Aut}.
 
\begin{table}[h]\footnotesize
  \caption{Controller Authentication Protocol}
  \label{tab:RAC}
  \begin{tabular}{|lll|} \hline 
    1. & $c \rightarrow AS$  & - Sends {\it controller's ID}, $Cr_c.ID_c$, to $KC$.\\
    2. & $AS$ &\\
       &2.1.& - Use $Cr_c.ID_c$ to check if $c$ is listed in\\ & & $KC$'s database.\\
       &2.2.& - Create a time stamp $TS_c$ and a session key\\ & & $S_{K_{TGS}}$ for the ticket.\\
       &2.3.& - Generate $T_c = Enc_{PK_{TGS}}(ID_c, TS_c, S_{K_{TGS}})$.\\
       &2.4.& - Retrieve $Cr_c.p_c$ from $KC$'s database.\\
       &2.5.& - Generate controller's secret key\\ & & $K_c = hash(Cr_c.p_c)$.\\
       &2.6.& - Create $M_A = Enc_{K_c}(S_{K_{TGS}})$,\\ & & $M_B = T_c$.\\
    3. & $AS \rightarrow c$ & $M_A$, $M_B$.\\
    4. & $c$ &\\
       &4.1.& - Generate a secret key $K_c = hash(Cr_c.p_c)$\\ & & from $c$'s password.\\
       &4.2.&	- Decrypt $M_A$ by $K_c$ to obtain $S_{K_{TGS}}$.\\
       &4.3.& - Create $M_C = [M_B, ID_{DBS}, Req_{DBS}]$\\ & & and $M_D = Enc_{S_{K_{TGS}}}(ID_c, TS_c)$.\\
    5. & $c \rightarrow TGS$ & - Sends $M_C, M_D$.\\
    6. & $TGS$ &\\
       &6.1.& - Decrypt $M_B$ in $M_C$ by $PK_{TGS}$ to obtain\\ & & $S_{K_{TGS}}$ and $ID_c$.\\
       &6.2.& - Decrypt $M_D$ by $S_{K_{TGS}}$ to get $ID_c$.\\
       &6.3.& - Compare $ID_c$ retrieved in step 6.1. and $ID_c$ \\ & & retrieved in step 6.2.
                If the result is "equal",\\ & & $c$ is authenticated to $TGS$.\\
    \hline
  \end{tabular}
\end{table}

\subsection{Controller Authorisation}
\label{sec:controller_authorization}
Two  kinds of  permissions were  investigated:  {\it Read}  and {\it  Write}. {\it  Write}
permission indicates the act of inserting a  new data into the  database. {\it  Write (W)}
permission cannot delete or change data that is already  stored in the database. {\it Read 
 (R)} permission is the ability to  retrieve  data  from the  database without  making any 
changes to the data.

\begin{example}
\label{ex:controllerAuthorization} 
Mobile App controller  can access the Cassandra original database  in the first horizontal
layer  (cf.  Figure  \ref{fig:architecture}),  but  it  cannot  access  the  de-identified
statistics  data.  Whereas, analysis  services  can  access the  de-identified anonymised database, but they cannot write data into the original mongoDB database. 
\end{example}

\begin{definition}[DAC Authority]
\label{def:DAC_Authority} 
Let  $CT=\{Ct_1, Ct_2,  \cdots, Ct_n\}$  be  the set  of  controllers in  the system.  Let
$CL=\{Cl_1, Cl_2,\cdots, Cl_n\}$ be the set of user collections in the system. Let $P=\{R,
W\}$  be the  set permissions  in the  system.  The set  of permissions  granted for  each
controller is defined as a 2-dimension array $DAC$. 
$$DAC = \{DAC_{(Ct_1,Cl_1)}, DAC_{(Ct_1, Cl_2)},DAC_{(Ct_1, Cl_3)}\cdots\}$$
where $$DAC_{(Ct_1, Cl_1)} = (Ct_1, Cl_1, P_{(Ct_1, Cl_1)})$$
$$DAC_{(Ct_1, Cl_2)} = (Ct_1, Cl_2, P_{(Ct_1, Cl_2)})$$
$$\cdots$$
\end{definition}

\begin{example} 
According to Figure \ref{fig:DAC}, we have $Ct = \{portal , mobile ,
analysis\  backend, \cdots\}$,  $Cl  =  \{clinic,  school, children,  \cdots\}$,
$P=\{R,W\}$.  A  DAC  authority  is  $DAC =  \{\{portal,  clinic,  \{R,W\}\},
\{portal, school, \{R, W\}\}, \cdots\}$. 
\end{example}

We use  LDAP protocol for communication  between the authorisation process  and the active
directory  (AD)  which  stores an  access  control  list  (ACL)  of permissions  for  each
controller.  The DAC  table (c.f.  Definition  \ref{def:DAC_Authority}) is  stored in  the
AD.  Each controller's  permissions are  defined in  an ACL  entry. Whereas,  the Kerberos
protocol authorises requests from the controller.  The LDAP active directory can be stored
in a separated  LDAP server or in  the same Kerberos server. The  authorisation process is
executed after the controller authentication was successful. 

The    authorisation   protocol    must   follow    the   authentication    process   (see
\ref{sec:Cont_Aut}). The  authorisation protocol is described  in steps 4, 5,  6 of Figure
\ref{fig:RAC}  and presented  in Table  \ref{tab:Kerberos_Controller_AUthorization}.  Once
$TGS$ compares the two  identities of the controller $c$, $TGS$ searches in  its AD for an
entry that defines  access permissions for $c$  to the database service  $DBS$. Then $TGS$
compares the access  permissions in $DAC_{(c,DBS)}$ and the request  $Req_{DBS}$.  If they
are  not equal,  $TGS$ sends  a reject  message to  $c$. Otherwise,  $TGS$ issues  two new
messages: $M_E$ and  $M_F$ (steps 1.1, 1.2) where  $M_E$ is the issued ticket  used by $C$
with  the service  $DBS$.  $M_E  =  Enc_{PK_{DBS}}(ID_c, TS_{c-DBS},  S_{K_{DBS}})$ is  an
encryption  made by  the  secrete key  of  service (i.e.,  $PK_{DBS}$),  of controller  ID
(i.e. $ID_c$), time  stamp (i.e. $TS_{c-DBS}$), and  the shared key of  the service (i.e.,
$S_{K_{DBS}}$); whereas,  $M_F = Enc_{S_{K_{TGS}}}(S_{K_{DBS}})$  is an encryption  of the
shared  key   of  the   service  (i.e.,   $S_{K_{DBS}}$)  made   by  $TGS$'s   shared  key
$S_{K_{DBS}}$. Then,  $TGS$ sends  $M_E$ and  $M_F$ to  the controller  $c$ (step  2). The
controller $c$ decrypts  $M_F$ with $TGS$'s shared key $S_{K_{TGS}}$  to obtain the shared
key of service $DBS$  (i.e., $S_{K_{DBS}}$), the time stamp of  controller $c$ and service
$s$  (i.e.,  $TS_{c-DBS}$),  and  the  identity of  controller  $c$  (i.e,  $ID_c$)  (step
3.1). Then,  $c$ creates a new  message $M_G$, that  is an encryption of  the controller's
identity (i.e.,  $ID_c$) and  the time stamp  of controller $c$  and service  $DBS$ (i.e.,
$TS_{c-DBS}$) (step 3.2). After that, controller $c$ sends $M_E$ and $M_G$ to the database
service $DBS$ (step 4).  When $DBS$ receives  the two messages from $c$, it first decrypts
$M_E$ using the secrete key of $s$ (i.e., $P_{K_{DBS}}$) to obtain the shared key of $DBS$
and the controller's identity $ID_c$ (step 5.1). $DBS$ decrypts $M_G$ using the shared key
$S_{K_{DBS}}$ to obtain  the controller's identity (i.e., $ID_c$) (step  5.2). Then, $DBS$
compares the two messages $M_E$ and $M_G$ (step 5.3), if the comparison results in "equal"
(i.e.,   $M_E=M_G$),   $c$    is   authorised   and   can   access    the   data   service
$DBS$. Simultaneously,  $DBS$ creates a  message $M_H$ that is  an encryption of  the time
stamp $TS_{c-DBS}$ granted to be used between $c$ and $DBS$. After that, $DBS$ sends $M_H$
to $c$  (step 6).  After  receiving $M_H$, $c$ decrypts  $M_H$ using service's  shared key
(i.e., $S_{K_{DBS}}$) to  obtain $TS_{c-DBS}$ (step 7.1). $c$ compares  the two time stamp
in the two messages retrieved from decrypting $M_F$ (step 3.1) and $M_H$ (step 7.1).
Finally, $c$ sequentially sends its requests to $DBS$. 

\begin{table}[h]\footnotesize
  \caption{Controller Authorisation Protocol}
  \label{tab:Kerberos_Controller_AUthorization}
    \begin{tabular}{|lll|} \hline 
         1.& & {\bf TGS}\\
         &1.1.& Search in AD for $c$'s entry $DAC_{(c,DBS)}$ .\\
         &1.2.& Compare access permissions in $DAC_{(c,DBS)}$ and\\&& $req_{DBS}$ in $M_C$.\\
         &1.3.& If equal, do step 1.3. Otherwise, sends a rejection\\&& response to $c$.\\
         &1.4.& Create\\&& $M_E=Enc_{PK_s}(ID_c, TS_{c-DBS}, S_{K_{DBS}})$.\\
         &1.5.& Create $M_F = Enc_{S_{K_{TGS}}}(S_{K_{DBS}})$.\\
         2.& & {\bf TGS} $ \rightarrow c \quad $  - Sends $M_E, M_F$.\\
         3.& & {\bf c}\\
         &3.1.& Decrypt $M_F$ by $S_{K_{TGS}}$ to obtain $S_{K_{DBS}}$, \\&& $TS_{c-DBS}$, and $ID_c$.\\
         &3.2.& Create $M_G = Enc_{S_{K_{DBS}}}(ID_c, TS_{c-DBS})$.\\
         4.& & $c \rightarrow$ {\bf DBS} $\quad $ - Sends $M_E, M_G$.\\
         5.& & {\bf DBS}\\
         &5.1.& Decrypt $M_E$ by $P_{K_{DBS}}$ to obtain $S_{K_{DBS}}$\\&& and $ID_c$.\\
         &5.2.& Decrypt $M_G$ by $S_{K_{DBS}}$ to obtain $ID_c$.\\
         &5.3.& Compare $M_E$ and $M_G$. If $M_E=M_G$, has $c$ \\ && authorised, and creates \\&&$M_H =          Enc_{S_{K_{DBS}}}(TS_{c-DBS})$.\\
         6.&& {\bf DBS} $\rightarrow c \quad $ - Sends $M_H$.\\
         7.& & {\bf c} \\
         &7.1.& Decrypt $M_H$ by $S_{K_{DBS}}$ to obtain $TS_{c-DBS}$.\\
         &7.2.& Compare $TS_{c-DBS}$ from steps 3.1 and 7.1.\\&& If $"="$, $DBS$ is authorised. \\
         
 		\hline
\end{tabular}
\end{table}

\section{Secure Data Transmission}
\label{sec:SDT}
When a subject, such  as a user or a controller, attempts to  access data at the back-end,
it has  to go through  two substantial security  shields, that is,  authentication process
(cf.    Section   \ref{sec:UserAUTH})    and    authorisation    process   (cf.    Section
\ref{sec:CAE}). Even though  such a data access  is rigorous, the risk of  data leakage is
still likely to occur, especially when storing and transmitting the data. Moreover, in our
system, the  data storage is distributed.  the databases are located  on different servers
(cf. Section \ref{sec:Archi}).  Any request to access the data the database servers
collaborate to output a response.

To   protect  data   storage  against   attacks,  the   storage  system   uses  encryption
algorithms. MongoDB supports various  encryption schemas\footnote{Securing MongoDB Part 3:
  Database                    Auditing                   and                    Encryption
  https://www.mongodb.com/blog/post/securing-mongodb-part-3-database-auditing-and-encryption},
such a default with  AES-256 in CBC and GCM mode.\footnote{Announcing the Advanced Encryption Standard (AES), NIST, FIPS 197, 2001} The  encryption schema can be
configured to comply with FIPS 140-2.\footnote{Security Requirements for Cryptographic Modules, NIST, FIPS 142, 2001} Cassandra supports Transparent Data
Encryption (TDE)\footnote{Oracle Advanced Security Transparent Data Encryption Best Practices, https://www.oracle.com/technetwork/database/security/twp-transparent-data-encryption-bes-130696.pdf}  for a lightweight encryption  of data and log  files that are
stored  in the  master  database of  server  that contains  administrative  data used  for
monitoring and controlling the system. 

Data transmission  also needs to  be protected  against eavesdroppers. Usually  the stored
data are queried and  transmitted in a plain form that can be  read in-flight by the third
party. We  adopted Secure Sockets Layer  (SSL)/ Transport Layer Security  (TLS) \cite{SSL}
for securing the data  transmitted over the network. This protocol  is applied between any
two components of  the system.  For example, as in  Figure \ref{fig:architecture}, the two
parties  needing a secure  data  transmission  between  themselves  included {\it Original 
MongoDB Database} and the module  {\it De-identification}, since  data  from {\it Original
MongoDB Database} is  transferred to  the module {\it De-identification}.   

\begin{table}[h]\footnotesize
  \caption{Secure Data Transmission Protocol}
  \label{tab:secure_data_transmission}
  \begin{tabular}{|lll|} \hline 
    1. & & $c \rightarrow s\quad $ Send\\
       & 1.1. & - SSL session request.\\
       & 1.2. & - Supported protocol version, list of cipher suits $CS$.\\
    2. & & {\bf s} \\
       & 2.1. & Create the request accepted.\\
       & 2.2. & Check supported protocol version.\\
       & 2.3. & Select cipher algorithms from $CS$.\\
    3. & & $s \rightarrow c\quad $ Send\\
       & 3.1. & - Selected cipher algorithm list.\\
       & 3.2. & - SSL certificate of the public key $K_s$.\\
       & 3.3. & - $s$'s public key $K_s$.\\
    4. & & {\bf c}\\
       & 4.1. & - Create session key $SK_{c-s}$.\\
       & 4.2. & - Encrypt $SK_{c-s}$ with $K_s$.\\
    5. & & $c \rightarrow s \quad $ Send $Enc_{P_s}(SK_{c-s})$.\\
    6. & & {\bf s} Decrypt $Enc_{P_s}(SK_{c-s})$ by private key $PK_s$ to get\\
       & & $SK_{c-s}$.\\
    \hline
  \end{tabular}
\end{table}

The goal of this  protocol is to secure the data while it  is exchanged between two nodes.
A    secure   protocol    using   SSL/TLS    is    presented   in    details   in    Table
\ref{tab:secure_data_transmission} between {\bf  c} and {\bf s}.  It is  assumed that {\bf
  c} and  {\bf s}  have their own  pair of asymmetric  keys including  a public key  and a
private key. We assume that all transactions made  between {\bf c} and {\bf s} are secured
by the session key, $SK_{c-s}$. Basically,  SSL/TLS connection is established based on the
handshake between the two sides {\bf c} and {\bf s}.  It is assumed that each side has its
own pair of public key and private key.   Whenever one side, e.g.  {\bf c}, has an SSL/TLS
connection request, {\bf c}  sends to {\bf s} a connection  request, its supported SSL/TLS
protocol version (e.g.,  TLS 1.2), and a list  of cipher suits that it  supports (step 1).
Cipher suit is a set of cryptographic  algorithms that {\bf c} supports, such as symmetric
(AES-256 CBC/GCM,  etc.) or asymmetric algorithms  (e.g., RSA), hash function  (e.g., SHA,
DES)  and  key  exchange  protocols  (e.g.    DHE),  etc.   An  example  of  cipher  suit:
{\it"ECDHE-RSA-AES128-GCM-SHA256"}. Once {\bf s} accepts  a connection request, it creates
an acceptance notification  (step 2.1). Then {\bf  s} checks if it also  supports the {\bf
  c}'s protocol version (step 2.2). If this is  the case, {\bf s} matches the cipher suits
in the list received from {\bf c} with its supported cryptographic algorithms, and selects
the suitable ones (step 2.3). {\bf s} sends  to {\bf c} the list of selected cryptographic
algorithms (step  3.1), an SSL  certificate (step 3.2), and  its public key  (i.e., $K_s$)
(step 3.3).  After receiving certificate and $K_s$ from {\bf s}, {\bf c} creates a session
key, $SK_{c-s}$ used for  all communications between {\bf c} and {\bf  s} (step 4.1). {\bf
  c} encrypts  the session key with  the public key $K_s$  to output $Enc_{P_s}(SK_{c-s})$
(step 4.2).   Then, {\bf c}  sends the encryption  $Enc_{P_s}(SK_{c-s})$ to {\bf  s} (step
5).  {\bf s}  uses its  private key  $PK_s$  to obtain  the session  key $SK_{c-s}$  (step
6). Then, the session key is used for  all transactions between {\bf c} and {\bf s}.  This
protocol is applied  for all transactions between any  of two modules or two  nodes in the
system.
 
\section{Evaluation}
\label{sec:SecEval}
In  this section,  we  evaluate the  proposed  model through  the  immune ability  against
attacks, that is, replay attack, eavesdropping attack and unauthorised spy. 

\paragraph{Replay attack} Usually  the adversary tries to get  packets transmitted through
the  network,  and  reuses them  later  in  the  following  sessions for  the  purpose  of
impersonating the sending  nodes. If the attack is successful,  the adversary receives the
response. In case  the response may contain sensitive information,  the adversary can take
advantage of the information for, which affects very badly the users privacy and sensitive
data leakage.  

\begin{lemma}
In the  proposed model, the  adversaries cannot reuse the  released packet for  the replay
attack purpose. 
\end{lemma}

\begin{proof}
In our  model, each  packet is stamped  with a  validation time. Based  on this  time, the
receiving nodes in the network check the  packet's expiration, by comparing its time stamp
with the  current time  to see if  the time  interval is greater  than the  requested time
threshold. If  this is  a case, the  packet is dropped.  The problem  is that how  long is
enough for  a packet's time to  live so that the  system can prevent the  replay attack in
time. 
\end{proof}
\paragraph{Eavesdropping  attack} The  eavesdropper tries  to stalk  packets found  in the
network  for the  further purposes.  They may  be benign  or malicious.  If they  are just
honest-but-curious  adversaries, they  read  the sensitive  information  in the  delivered
packet  only to  satisfy their  curiosity.  However, this  type  of attack  can leave  the
backdoor in the system.  Then, the more malicious adversary may  be, successfully do their
attack will be. With malicious adversaries, they are more harassed and dangerous. They aim
to get  more personal  information to  get more  benefits, and  may seriously  harm user's
honour, privacy and finance, or to interfere with the data to change data integrity.  

\begin{lemma}
in  the  proposed model,  Eavesdroppers  cannot  modify the  package  content  or get  any
sensitive information from the communication tunnel. 
\end{lemma}

\begin{proof}
Hence, in  our proposed model, we  protect user data  as they are transferred  through the
network by adopting  the SSL/TLS protocol. This  protocol creates a robust  tunnel to hide
all transmitted  data especially  encrypting them  right when they  are released  from the
application and  at the transport layers.  Moreover, with SSL/TLS protocol,  a cipher suit
exploits different encryption algorithms and  hash functions. So, eavesdroppers cannot get
any sensitive information from the tunnel.  
\end{proof}

\paragraph{Unauthorised  spy} Our  model can  protect data  against unauthorised  spies by
authenticating and authorising the users and  the controllers attempting to access data at
the back-end.  Spies can  abuse the  lack of  security check,  such as  authentication and
authorisation, to steal  important information. After retrieving  the crucial information,
they turn into the previous eavesdropping adversaries, and get either benign or malicious.
\begin{lemma}
In the  proposed model, the user  and controller cannot  access database and use  the data
with any purpose. 
\end{lemma}

\begin{proof}
  In the proposed model,  the granted permissions of a user and  controller must be stored
  in the active directory centre to be able to access database.  When a user or controller
  sends a data access request, the server will check for suitable permission in the active
  directory. If  this is a case,  the server passes  the request to the  back-end database
  server.
\end{proof}

\section{Experimental Results}
\label{sec:Experi}


This research  is a part of  a larger project which  focuses on monitoring the  changes of
obesity-related behavioural  risk factors (what and  how children eat, how  they move, how
they sleep) to the  prevalence of obesity, and to define  a parsimonious behavioural model
and data structure that will be free of redundant individual information and will minimise
the use of sensitive  information. The users of our system  include the clinicians, school
teachers, and public health authorities.

To achieve  such objectives, first  we build a software  system that collects,  stores and
analyses  the  data sets of children and adolescent  from  different locations  around
Europe.  The children  and  patients join the  system on  voluntarily  basis. 

The  system
collects  their  behaviours   and  relevant information (e.g., daily exercises,  food
consumption, junk  food quantity, locations, clinics, schools, groups, etc.),  and stores
them in the two major databases: MongoDB and Cassandra. MongoDB is the core database where
the most  of data is stored.  Cassandra is used  for collecting the time-series  data from
users' smart  devices which are processed  then stored in  the local storage of  the smart
devices before they are transferred to MongoDB (c.f. Section \ref{sec:database_schema}).

To  provide clinicians  and  developers a  tool  to  access the  above  two databases,  we
implement a  set of  interfaces: Clinical  Advisor (CA) services.  A description  of these
services are presented in Appendix \ref{sec:AppendixA}.  Each service gives an output data
through  a graph  which the  clinicians  can use  for comparing  situations of  individual
children  or groups  of  children, to  monitor  the children  patients,  to recommend  the
effective treatments or needed therapies suitable to each patient.

Each user  in the system  must be granted permissions.  User permissions are  discussed in
Section \ref{sec:UserAUTH}. Moreover,  these services can be accessed  by other components
in the system, such as controllers or modules (c.f. Section \ref{sec:CAE}).

\subsection{Database Schema}
\label{sec:database_schema}
\begin{figure*}
      \centering    
\includegraphics[height=13cm, width=16cm]{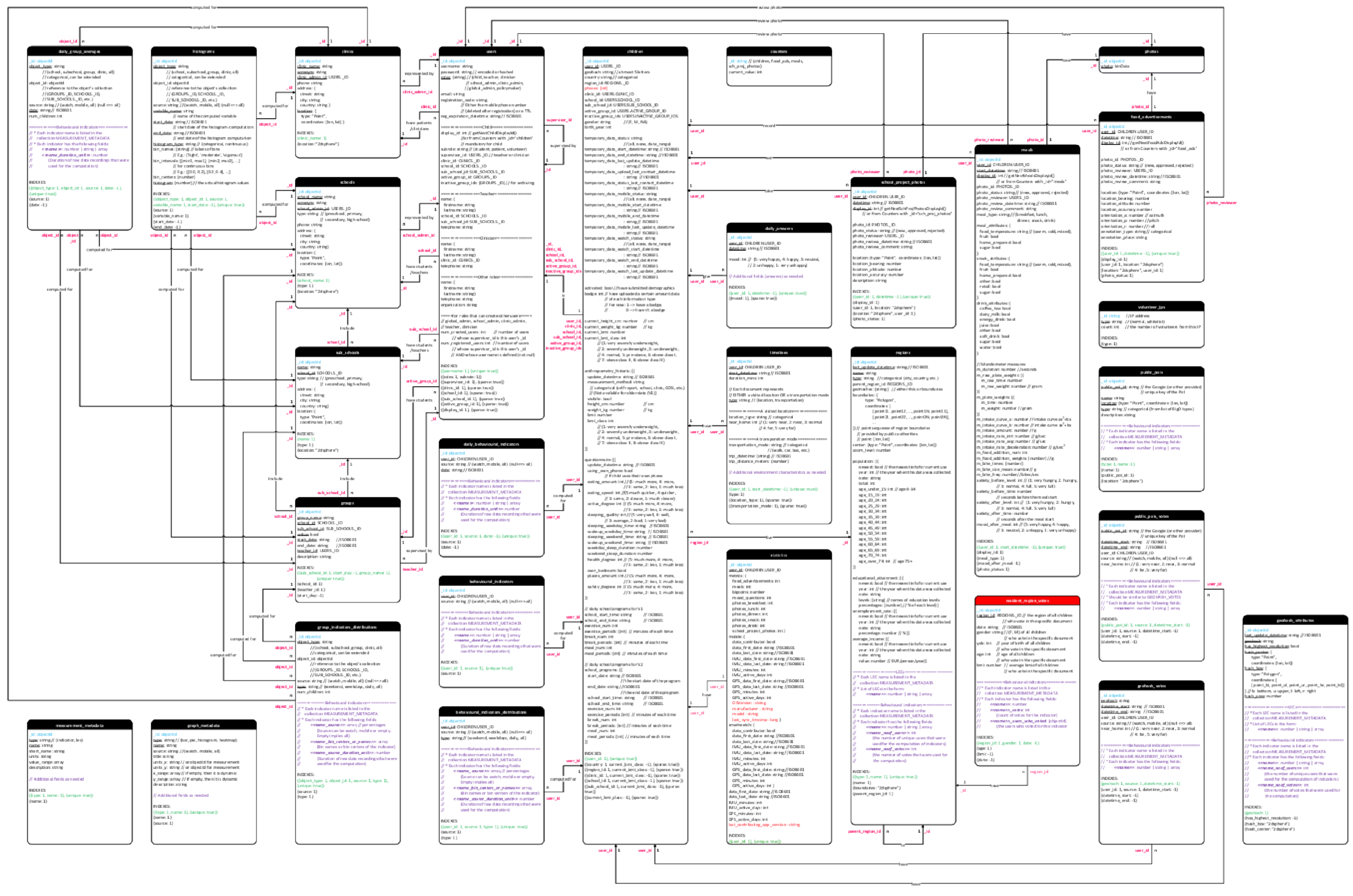}
\caption{MongoDB schema of our system}\label{fig:MongoDB_schema}

\centering
\includegraphics[height=1.2cm, width=12cm]{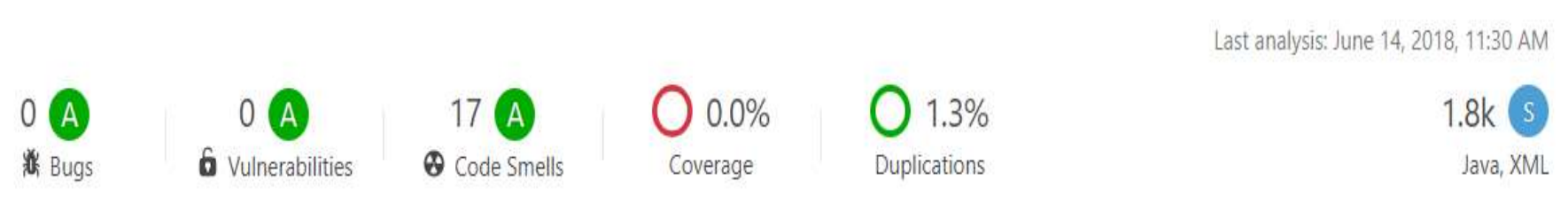}
\caption{Code Quality Test}\label{fig:UCD_Test}

    \end{figure*}

\begin{figure}[h]
\centering
\includegraphics[height=7cm, width=7cm]{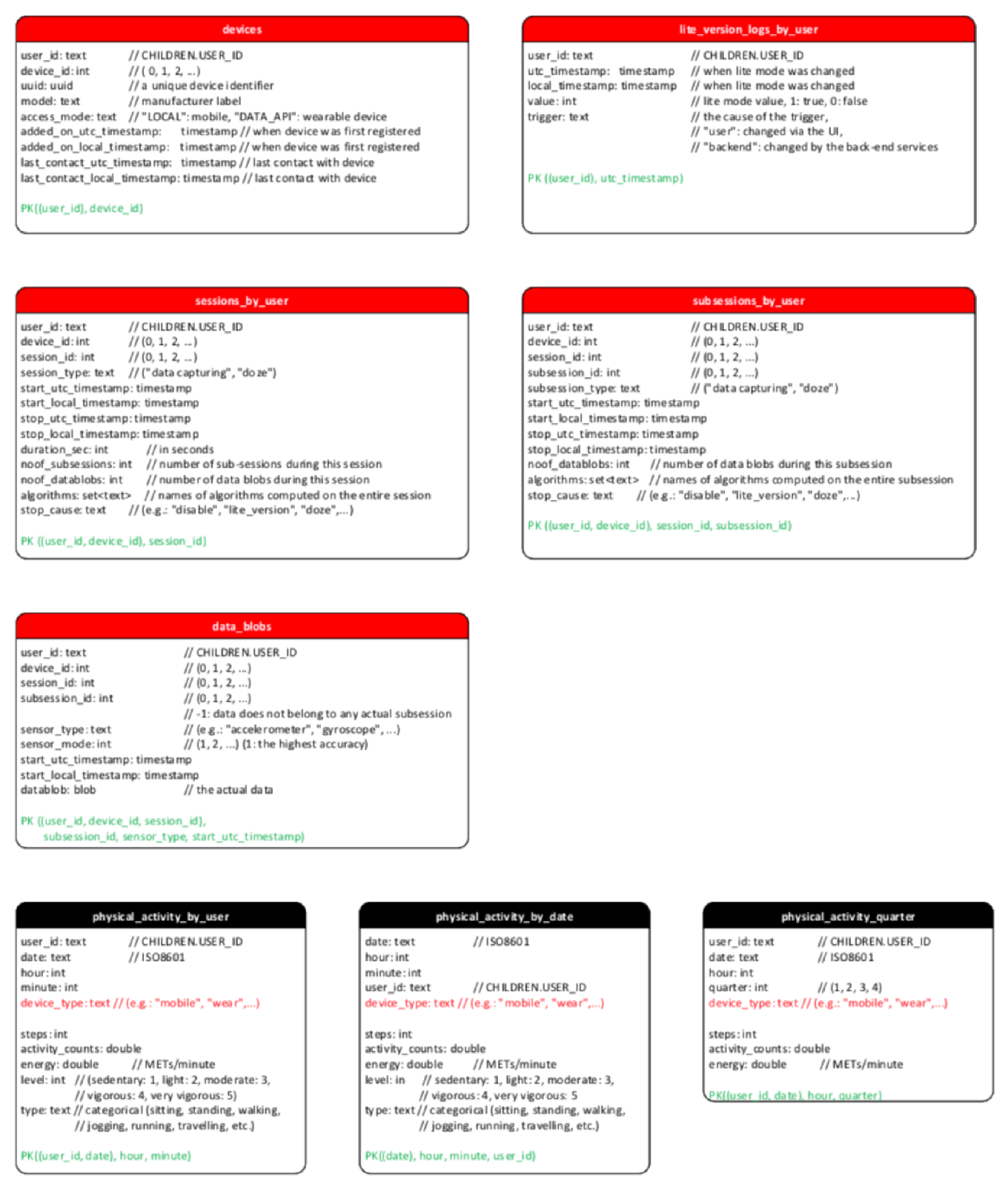}
\caption{Cassandra Database Schema}\label{fig:Cassandra_schema}
\end{figure}

In order to  provide a full pack of  Clinical Advisor services, we design  and implement a
database in  term of $28$  collections in MongoDB (c.f.  Figure \ref{fig:MongoDB_schema}),
such as  regions, groups, children, students,  etc., and $8$ tables  in Cassandra database
(c.f.  Figure \ref{fig:Cassandra_schema}),  such as  devices, sessions  by user,  physical
activities by user, physical activities by date, etc. 

\subsection{Deployment Environment}
\label{sec:exp__dataset_environment}

In the proposed software  system, there are two main modules,  that is, Application module
and Analysis module. Each  module includes a front-end server and a  cluster of nodes. The
Application module is responsible for data  storage in Cassandra and MongoDB, receives the
requests, and provides services for Mobile App Controller, Web App Controller, and Portals
Controller. This module hosts an Apache Balancer and a Tomcat Server. TLS/SSL and Kerberos
are set  up in  this server. Whereas,  the Analysis module  provides the  computations for
analysis services.  This module  also provides  the interfaces  for Back-end  Analysis and
Services. Currently, they are all deployed on  the same server. For our implementation and
testing, the system uses Apache Server  $2.4$, Tomcat $8$, Cassandra $>3.1$, MongoDB $3.4$
(Community       version),       Java,       Python,       JSON,       Flask,       Apache
Maven\footnote{https://maven.apache.org/},  JUnit\footnote{https://junit.org/junit5/}, and
SonarQube\footnote{https://www.sonarqube.org/}. The Tomcat server contains  the REST APIs as interfaces of the  system for controllers and
end-users.  They contain  different  APIs  depending on  the  required functionality.  The
Keberos and  LDAP servers are always  connected to the  Tomcat server to get  the requests
from users and controllers, then process the request and reply back to the Tomcat server.


\subsection{System Quality Test}
\label{sec:system_test}
We test the Clinical Advisor services after  they were integrated into the main system. We
run various types of tests.  For code  quality testing, we use the software SonarQube, and
for code  testing we  use the  tool JUnit.  SonarQube measures  the following  criteria to
ensure that the  system has high quality  code; no bugs, no  vulnerabilities, no technical
debts, no code duplication, and no duplicated blocks. 

Figure \ref{fig:UCD_Test} presents the testing results of the whole system after executing
the unit testing  and the code quality testing.  The code smells are pieces  of code that,
while technically not wrong, should be altered. As all the returned measurements are rated
A that means the code is of high quality. 

\section{Conclusion}
\label{sec:Concl}
This paper,  we show  that designing  and implementing  a system  for large  and sensitive
data-driven application is not straightforward, mainly  the storage system, where the data
should be stored and preserved from any unauthorised access. Moreover, this is not enough,
as the data needs to be accessed for analysis purposes. This creates another challenge, as
it has to  be protected during its transit  over the network. We can argue  that these are
provided by the operating system of both clients and hosts machines. However, we showed in
this  paper  that  this is  not  enough.  Sensitive  data  should and  must  be  protected
independently  of the  underlying operating  system or  the system  security put  in place
either at client or host.  

The proposed architecture not only incorporates security and protection by design but also
provides  a   framework  for   any  system  architecture   that  deals   with  data-driven
applications. Our architecture separates between  access permissions to original data, and
its derived data for special use (statistics and analysis for instance). We also provide a
set of services for  the external users to access the back-end. We  provide a formal model
to prove the system resilience to attacks and implemented it using popular and tools.

As  a future  work, one  needs  to provide  an  automated security-aware  system for  data
analysis. We will to generalise the proposed  architecture to any system where the data is
the main focus.  We will also improve  the anonymised data in the architecture (see Figure
\ref{fig:architecture}), to  provide a set of  secure interfaces for external  users to do
data mining on the anonymous data.

\appendices
\section{Clinical Advisor (CA) Graphs}
\label{sec:AppendixA}
We implemented the graphs to support clinicians in observing and helping the patients. The
Clinical Advisor module of the system provides  basic statistics  via
two RESTful web services interface, \textit{Individual} and \textit{Population}. 

\subsection{Individual Services}
\label{sec:individual}
An {\it indicator} is measurable quantity to provide the information about an individual’s
behaviour in eating, physical activity and sleep. This service provides statistical graphs
of the  user. Given a time  period and an indicator,  these graphs show how  the indicator
changes during that period.

\subsubsection{Numeric Indicators}
\label{sec:numeric_indicator}
There are  three types of  graphs for Individual  services for supporting  clinicians. The
changes are displayed by average values.
\begin{figure}[h]
\centering
  \includegraphics[height=5cm, width=6cm]{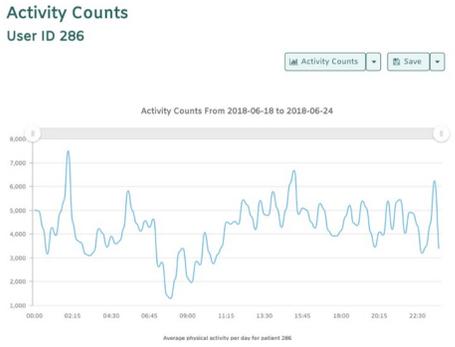}
  \caption{A graph of activity counts in Clinical Portal.}
  \label{fig:account_activity}
\end{figure}

\begin{itemize}
   \item {\it Daily MET.} This graph shows the average MET (Metabolic equivalent of task),
         which  expresses the  energy  cost  in physical  activities  during  a period  of
         time.  Higher  MET  corresponds  to   heavier  activities.  This  graph  supports
         clinicians in checking the effectiveness of physical activities.
   \item {\it Daily  physical activity  counts.} This  graph shows  the average  activity
         counts  during a  given time  period (Figure  \ref{fig:account_activity}). Higher
         counts correspond to higher physical activities.  This will 
         support clinicians in observing the effectiveness of physical activities. 
   \item {\it Daily physical activity steps.}  This graph shows the average activity steps
         during a  time period. This  graph provides clinicians  a tool of  monitoring the
         most popular types of physical activities (walking, running, etc.).
\end{itemize}

\subsubsection{Categorical Indicators}
\label{sec:categorical_indicator}
Categorical indicators  allow to  provide graphs  for meal  types, visited  locations, and
transportation modes.  Changes are expressed by the distribution of values. 

\begin{itemize}
   \item {\it  Daily dietary.}  This graph  shows the distribution  of meals  at different
         times during the  day. With this graph, clinicians can  get information about the
         dietary habits of children (c.f. Figure \ref{fig:daily_meal_child}).
         \begin{figure}[h]
           \vspace{-1cm}
           \hspace{1cm}
           \includegraphics[height=7cm, width=8cm]{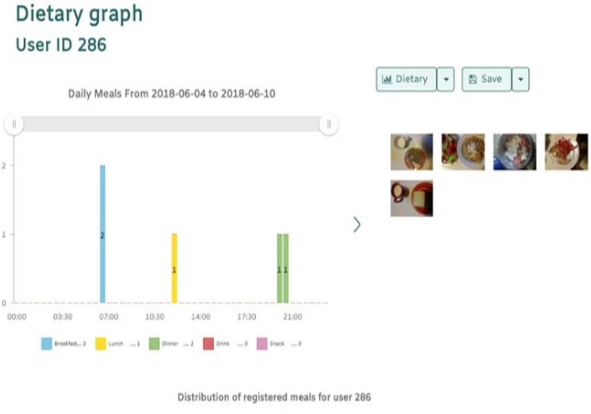}
           \vspace{-1cm}
           \caption{A graph of daily meals of a child in Clinical Portal}
           \label{fig:daily_meal_child}
         \end{figure} 
   \item  {\it Daily  visited locations.}  This  graph shows  the location  types a  child
         usually  visits. With  this  graph,  clinicians can  learn  more about  favourite
         locations of children and their habits. For example, if children prefer sedentary
         activities or dynamic activities, how often children join community activities or
         use recreational facilities.
   \item {\it Daily transportation.}  This graph presents commuting travels and frequent
        travelling methods.  Clinicians know their travel habits.
\end{itemize}

\subsection{Population Services}
\label{sec:population}
\begin{figure}[h]
\centering
  \includegraphics[height=4cm, width=6cm]{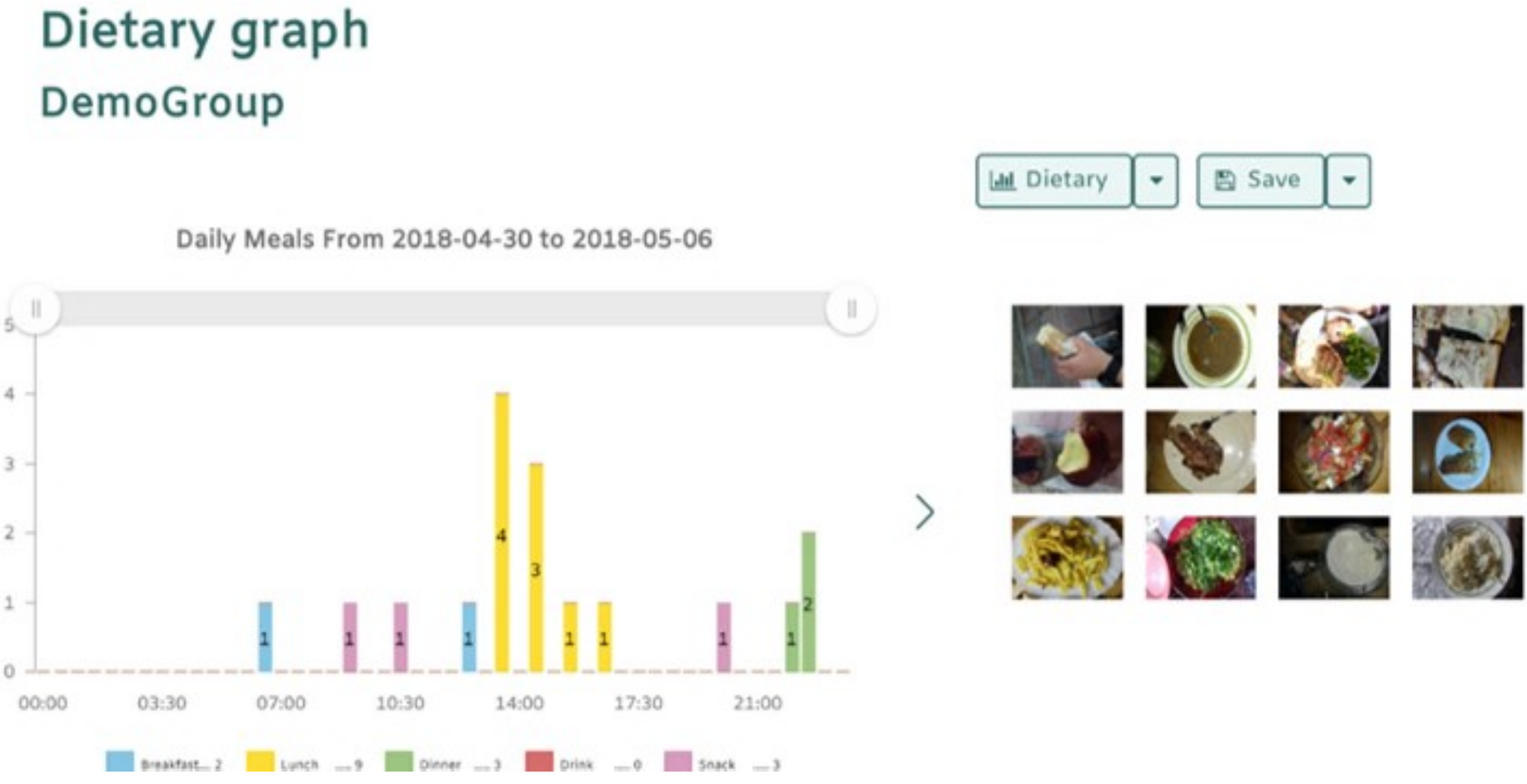}
  \caption{A graph of daily meals of a group in Clinical Portal}\label{fig:daily_meal_group}
\end{figure} 

A population  can be patients  of a  clinic, students of  a school. The  population graphs
provide the  statistical data of a  population. These graphs contain  the body measurement
distributions  (height, weight,  BMI class)  that are  typical indicators  to consider  to
measure  children body  fat. Using  these graphs,  clinicians can  know the  prevalence of
obesity in a community or population. 

\subsection{Mock-up Services}
As a future works, we will implement more graphs to support clinicians. The planned graphs
are listed below: 
\begin{itemize}
  \item  {\it  Graphs  for  food  consumption  and  body  measurements}  shows  the
        relationship between  food consumption and  body measurements (e.g.,  weight, BMI)
        over a time period. 
  \item {\it Graphs for physical activities  and body measurements} shows the relationship
        between physical activities and body measurements  (e.g., weight, BMI) over a time
        period. 
  \item {\it  Graphs for food  environment and  body measurements} shows  the relationship
        between food  environment and body  measurements (e.g.,  weight, BMI) over  a time
        period. 
\end{itemize}

\section*{Acknowledgment}
The work leading to these results has received funding from the European Community's Health, demographic change and well-being Programme under Grant Agreement No. 727688, 01/12/2016 - 30/11/2020

\bibliographystyle{plain}
\bibliography{References}

\begin{thebibliography}{10}

\bibitem{Ahmad2016}
M.~A. Ahmad.
\newblock Secfhir: A security specification model for fast healthcare
  interoperability resources.
\newblock {\em International Journal of Advanced Computer Science and
  Applications (IJACSA)}, 7(6), 2016.

\bibitem{Tahar2015}
Aadil Al-Mahrouqi, Sameh Abdalla, and Tahar Kechadi.
\newblock Cyberspace forensics readiness and security awareness model.
\newblock {\em International Journal of Advanced Computer Science and
  Applications}, 6(6), 2015.

\bibitem{Blobel2016}
B.~Blobel, D.~M. Lopez, and C.~Gonzalez.
\newblock Patient privacy and security concerns on big data for personalized
  medicine.
\newblock {\em Springer Journal in Health and Technology}, 6(1):75--81, 2016.

\bibitem{Boon2009}
A.~Boonyarattaphan, B.~Yan, and C.~Sam.
\newblock A security framework for e-health service authentication and e-health
  data transmission.
\newblock In {\em 9th International Symposium on Communications and Information
  Technology}, pages 1213--1218, Incheon, Korea, September 2009.

\bibitem{Barbara2013}
B.~Carminati, E.~Ferrari, and N.~H. Tran.
\newblock Enforcing trust preferences in mobile person-to-person payments.
\newblock In {\em 2013 International Conference on Social Computing}, pages
  429--434, Sep. 2013.

\bibitem{Barbara2014}
B.~Carminati, E.~Ferrari, and N.~H. Tran.
\newblock Secure web service composition with untrusted broker.
\newblock In {\em 2014 IEEE International Conference on Web Services}, pages
  137--144, June 2014.

\bibitem{Barbara2015}
B.~Carminati, E.~Ferrari, and N.~H. Tran.
\newblock A privacy-preserving framework for constrained choreographed service
  composition.
\newblock In {\em 2015 IEEE International Conference on Web Services}, pages
  297--304, 2015.

\bibitem{Rule15}
S.~Cohen and K.~Mitchell.
\newblock White paper: Wellness plans -- diagnosing compliance concerns.
\newblock {\em FYI - In Depth, Xerox}, 38, December 2015.

\bibitem{Dubo2017}
A.~Dubovitskaya, Z.~Xu, S.~Ryu, M.~Schumacher, and F.~Wang.
\newblock Secure and trustable electronic medical records sharing using
  blockchain.
\newblock {\em Annual Symposium proceedings (AMIA)}, pages 650--659, 2018.

\bibitem{dwi2019}
Ashutosh~Dhar Dwivedi, Gautam Srivastava, Shalini Dhar, and Rajani Singh.
\newblock A decentralized privacy-preserving healthcare blockchain for iot.
\newblock {\em Sensors}, 19(2), 2019.

\bibitem{Tahar2006}
Tariq~N. Ellahi, Benoit Hudzia, Liam McDermott, and M.~Tahar Kechadi.
\newblock Security framework for {P2P} based grid systems.
\newblock In {\em 5th International Symposium on Parallel and Distributed
  Computing (ISPDC)}, pages 230--237, Timisoara, Romania, July 2006.

\bibitem{Tahar2016}
Saad Fehis, Omar Nouali, and M-Tahar Kechadi.
\newblock {\em A New Distributed Chinese Wall Security Policy Model}, 11(4),
  2016.

\bibitem{RBAC}
D.~Ferraiolo and R.~Kuhn.
\newblock Role-based access controls.
\newblock In {\em 15th NIST-NCSC National Computer Security Conference}, pages
  554--563, Baltimore, MD, USA, March 2009.

\bibitem{Guo2012}
L.~Guo, C.~Zhang, J.~Sun, and Y.~Fang.
\newblock Paas: A privacy-preserving attribute-based authentication system for
  ehealth networks.
\newblock In {\em IEEE 32nd International Conference on Distributed Computing
  Systems}, pages 224--233, Macau, China, June 2012.

\bibitem{Huang2017}
H.~Huang, T.~Gong, N.~Ye, R.~Wang, and Y.~Dou.
\newblock Private and secured medical data transmission and analysis for
  wireless sensing healthcare system.
\newblock {\em IEEE (Journal) Transactions on Industrial Informatics},
  13(3):1227--1237, June 2017.

\bibitem{Kerberos}
J.~Kohl and C.~Neuman.
\newblock White paper: The kerberos network authentication service (v5).
\newblock \url{https://tools.ietf.org/html/rfc4120}, 2005.

\bibitem{Kotz16}
D.~Kotz, C.~A. Gunter, S.~Kumar, and J.~P. Weiner.
\newblock Privacy and security in mobile health: A research agenda.
\newblock {\em Journal of Computer}, 49(6):22--30, 2016.

\bibitem{Liu2015}
W.~Liu, X.~Liu, J.~Liu, Q.~Wu, J.~Zhang, and Y.~Li.
\newblock Auditing and revocation enabled role-based access control over
  outsourced private ehrs.
\newblock In {\em IEEE 7th International Symposium on Cyberspace Safety and
  Security}, pages 336--341, New York, USA, 2015.

\bibitem{Marci2006}
M.~Marci, R.~Tanya, and S.~Shankar.
\newblock Security and privacy issues with health care information technology.
\newblock In {\em 28th IEEE EMBS Annual International Conference}, pages
  5453--5458, New York, USA, 2006.

\bibitem{Marcos2015}
A.~Marcos, H.~Leonardo, M.~Bruno, C.~Tereza C.~M. B., and N.~Mats.
\newblock Secourhealth: A delay-tolerant security framework for mobile health
  data collection.
\newblock {\em IEEE Journal of Biomedical and Health Informatics},
  19(2):761--772, 2015.

\bibitem{Miao2016}
Yinbin Miao, Jianfeng Ma, Ximeng Liu, Fushan Wei, Zhiquan Liu, and Xu~An Wang.
\newblock m2-abks: Attribute-based multi-keyword search over encrypted personal
  health records in multi-owner setting.
\newblock {\em Journal of Medical Systems}, 40(11):246, Oct 2016.

\bibitem{ThienAnetal}
T.~A. Nguyen, N.~A. Le-Khac, and M.~T. Kechadi.
\newblock Privacy-aware data analysis middleware for data-driven ehr systems.
\newblock In {\em Future Data and Security Engineering (FDSE)}, pages 335--350,
  Ho Chi Minh City, Vietnam, 2017.

\bibitem{SSL}
R.~Oppliger.
\newblock {\em SSL and TLS: Theory and Practice}.
\newblock Information Security and Privacy. Artech House, 2005.
\newblock ISBN 978-1596934474.

\bibitem{Ostherr17}
K.~Ostherr, S.~Borodina, R.~C. Bracken, C.~Lotterman, E.~Storer, and
  B.~Williams.
\newblock Trust and privacy in the context of user-generated health data.
\newblock {\em Journal of Big Data and Society}, 4(1), 2017.

\bibitem{Rao15}
S.~Rao, S.~N. Suma, and M.~Sunitha.
\newblock Security solutions for big data analytics in healthcare.
\newblock In {\em Second International Conference on Advances in Computing and
  Communication Engineering}, pages 510--514, Dehradun, India, 2015.

\bibitem{Schatz15}
B.~R. Schatz.
\newblock National surveys of population health: Big data analytics for mobile
  health monitors.
\newblock {\em Journal of Big Data}, 3(4):219--229, 2015.

\bibitem{LDAP}
J.~Sermersheim.
\newblock Lightweight directory access protocol (ldap): The protocol, 2006.
\newblock Copyright (C) The Internet Society (2006).
  https://tools.ietf.org/html/rfc4511.

\bibitem{Shafer2010}
J.~Shafer, S.~Rixner, and A.~L. Cox.
\newblock The hadoop distributed filesystem: Balancing portability and
  performance.
\newblock In {\em IEEE International Symposium on Performance Analysis of
  Systems and Software (ISPASS)}, pages 122--133, White Plains, New York, 2010.

\bibitem{QI}
L.~Sweeney.
\newblock Project technical report: Simple demographics often identify people
  uniquely.
\newblock 2000.
\newblock http://dataprivacylab.org/projects/identifiability/paper1.pdf.

\bibitem{Yu15}
W.~D. Yu, C.~Pratiksha, S.~Swati, S.~Akhil, and M.~Sarath.
\newblock A modeling approach to big data based recommendation engine in modern
  health care environment.
\newblock In {\em 39th Annual Computer Software and Applications Conference},
  pages 75--86, Taichung, Taiwan, 2015.

\bibitem{Zhang2010}
R.~Zhang and L.~Liu.
\newblock Security models and requirements for healthcare application clouds.
\newblock In {\em IEEE 3rd International Conference on Cloud Computing}, pages
  268--275, Florida, USA, 2010.

\end{thebibliography}




\end{document}